\documentclass[sigconf]{acmart}

\usepackage{graphicx}
\usepackage{xcolor}
\usepackage{tikz}

\definecolor{clonecolor}{HTML}{0041CF}
\definecolor{extendcolor}{HTML}{15803D}
\definecolor{proxycolor}{HTML}{BE123C}
\definecolor{complementcolor}{HTML}{C2410C}
\definecolor{standbycolor}{HTML}{A16207}
\definecolor{partitioncolor}{HTML}{0F766E}

\newcommand{\circledstrategy}[2][clonecolor]{%
  \tikz[baseline=(label.base)]{
    \node[
      circle,
      fill=#1,
      text=white,
      draw=none,
      inner sep=0pt,
      minimum size=1.25em,
      font=\scriptsize\bfseries
    ] (label) {#2};
  }%
}

\newcommand{\circledstrategyinline}[2][clonecolor]{%
  \tikz[baseline=(label.base)]{
    \node[
      circle,
      fill=#1,
      text=white,
      draw=none,
      inner sep=0pt,
      minimum size=1.1em,
      font=\tiny\bfseries
    ] (label) {#2};
  }%
}

\newcommand{\designstrategy}[3][clonecolor]{%
  \par\addvspace{0.55\baselineskip}%
  \noindent
  \circledstrategy[#1]{#2}%
  \hspace{0.45em}%
  \textcolor{#1}{\textbf{#3.}}\enspace
}

\newcommand{\inlinestrategy}[3][clonecolor]{%
  \circledstrategyinline[#1]{#2}%
  \hspace{0.22em}%
  \textcolor{#1}{\textbf{#3}}%
}

\newcommand{\inlineicon}[1]{%
  \raisebox{-0.12em}{\includegraphics[height=.95em]{#1}}%
}

\newcommand{\iconUser}{\inlineicon{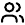}}
\newcommand{\iconMtm}{\inlineicon{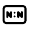}}
\AtBeginDocument{%
  }

\setcopyright{acmlicensed}
\copyrightyear{2018}
\acmYear{2018}
\acmDOI{XXXXXXX.XXXXXXX}
\acmConference[Conference acronym 'XX]{Make sure to enter the correct
  conference title from your rights confirmation email}{June 03--05,
  2018}{Woodstock, NY}
\acmISBN{978-1-4503-XXXX-X/2018/06}

\begin{document}

\title{MRxD: A Usage-Pattern Framework for Mixed-Reality Cross-Device Interaction}

\author{Yang Zhan}
\affiliation{%
  \institution{North Carolina State University}
  \city{Raleigh}
  \state{North Carolina}
  \country{USA}
}
\email{yzhan3@ncsu.edu}

\author{Yuxuan Huang}
\affiliation{%
  \institution{North Carolina State University}
  \city{Raleigh}
  \state{North Carolina}
  \country{USA}
}
\email{yhuang94@ncsu.edu}

\author{Yuchen Cai}
\affiliation{%
  \institution{McMaster University}
  \city{Hamilton}
  \state{Ontario}
  \country{Canada}
}
\email{caiy72@mcmaster.ca}

\author{Yichen Andy Yu}
\affiliation{%
  \institution{North Carolina State University}
  \city{Raleigh}
  \state{North Carolina}
  \country{USA}
}
\email{yyu55@ncsu.edu}

\author{Xiaoran Yang}
\affiliation{%
  \institution{North Carolina State University}
  \city{Raleigh}
  \state{North Carolina}
  \country{USA}
}
\email{xyang49@ncsu.edu}

\author{Ye Yuan}
\affiliation{%
  \institution{McMaster University}
  \city{Hamilton}
  \state{Ontario}
  \country{Canada}
}
\email{irene.yuan@mcmaster.ca}

\author{Qiao Jin}
\affiliation{%
  \institution{North Carolina State University}
  \city{Raleigh}
  \state{North Carolina}
  \country{USA}
}
\email{qjin4@ncsu.edu}

\renewcommand{\shortauthors}{Zhan et al.}

\begin{abstract}
Mixed reality (MR) cross-device experiences often combine MR headsets with non-immersive devices such as laptops, tablets, and phones. Yet such cross-device design remains challenging because users must coordinate workflows across heterogeneous devices with different roles. This motivates our focus on usage patterns: recurrent ways in which people distribute and coordinate work across devices. Through a systematic analysis of 238 MR cross-device papers, we developed MRxD, a usage-pattern framework organized around three dimensions: user attention, task relatedness, and cognitive symmetry. Using these dimensions, we identified six usage patterns and synthesized the design strategies commonly used within each pattern. We further investigated the framework's application in design practice through a workshop (n=6) and the results suggest that MRxD can guide pattern-based design exploration and identify blind spots. We further discuss how usage patterns may shift or combine, and outline research gaps and design implications for future MR cross-device systems.
\end{abstract}


\maketitle

\section{Introduction}

Mixed Reality (MR)\footnote{MR refers to the combination of Virtual Reality (VR) and Augmented Reality (AR) in this paper, consistent with broad definitions of MR in prior work~\cite{speicher2019mixed,skarbez2021revisiting}.} cross-device interaction combines head-mounted displays (HMDs) with non-immersive devices (e.g., phones, desktops) to support activities that are difficult to handle through the headset alone. While HMDs provide immersive and spatial presentation, non-immersive devices often remain better suited for dense reading~\cite{mrcd018}, long text entry~\cite{mrcd206}, and routine productivity tasks~\cite{mrcd004,mrcd181}. These complementary strengths have motivated prior systems to explore this combination in varied ways, including augmented displays~\cite{mrcd285,ads-survey}, control offloading~\cite{mrcd011,mrcd013, mrcd025,mrcd194}, and desktop-to-MR authoring~\cite{mrcd242,mrcd191}. MR cross-device interaction is more than adding peripheral devices to an HMD, but a broader design space for coordinating various display and interaction resources. 
Prior cross-device research describes this design space through the distribution of interfaces, information, and interaction across device ecologies~\cite{cd-taxonomy}. A recent work on MR Hybrid User Interfaces (MR-HUI) revisits the earlier notion of HUI for contemporary MR systems~\cite{mr-hui}. Building on Feiner's view~\cite{feiner-hui} of HUIs as combinations of heterogeneous display and interaction technologies, MR-HUI characterizes systems that combine conventional 2D devices with MR environments in complementary roles~\cite{mr-hui}. Their synthesis describes these systems mainly through device and component configurations, dependency, anchoring, interaction dynamics, and system-level challenges.

However, device and component configuration alone does not fully explain how users interact with the cross-device system. The same device configuration can support different forms of cross-device activities. For example, an HMD-phone setup may use the handheld device as a standard smartphone~\cite{mrcd194}, a menu surface~\cite{mrcd149}, a spatial controller~\cite{mrcd094}, or a text-entry device~\cite{mrcd063}. Despite the identical hardware setup, the underlying patterns of attention allocation, action flow, and cross-device interpretation beyond the phone's functional role may differ substantially, leading to distinct design requirements~\cite{rashid2012factors}. In cross-device research, usage-oriented analyses have been widely used to describe and form how cross-device systems fit into concrete activities in addition to how devices are connected~\cite{irene-usage,cd-education}, which helps explain why the same device ecology can require different design choices.

This motivates a usage-centric account of MR cross-device interaction. Existing system-oriented frameworks identify user-facing issues such as visual attention switching and perceptual synchronization~\cite{mr-hui} as vital challenges, but the usage patterns behind them remain under-specified. We use the term \textbf{usage pattern} to describe recurring ways in which users coordinate immersive and non-immersive devices during a usage activity~\cite{irene-usage}. A usage-pattern framework can support both analysis and design in MR cross-device interaction. It helps researchers compare systems by how devices work together in use beyond device setup. It also helps designers begin with intended user activities before deciding which devices and interaction techniques to use. 
Thus, this paper develops a usage-centered synthesis of MR cross-device interaction. We ask:

\begin{enumerate}
  \item[\textbf{RQ1.}] What usage patterns are reported in prior MR cross-device research?
  \item[\textbf{RQ2.}] What design strategies are associated with these usage patterns, and how can these design strategies be applied to generate new MR cross-device designs?
\end{enumerate}

To answer these questions, we conducted a systematic review of MR cross-device research published between January 2015 and December 2025. The review followed a PRISMA-informed~\cite{page2021prisma} collection and screening process and resulted in a corpus of 238 papers. We then developed a usage-pattern framework using a hybrid deductive--inductive method, applied it across the corpus, and inductively synthesized recurring design strategies within each pattern. Based on the findings, we proposed \textbf{MRxD}, a framework characterizing MR cross-device usage through three dimensions: \textit{User Attention}, or how users distribute attention across devices; \textit{Task Relatedness}, or whether devices support the same activity or separate activities; and \textit{Cognitive Symmetry}, or whether users can carry the same interpretation across devices or must construct a new interpretive mapping between them. We use these dimensions to identify recurring usage patterns and to organize design strategies that prior systems use to support them. A design workshop ($n=6$) was further conducted to examine how the framework can support design exploration. In summary, this paper contributes:

\begin{enumerate}
  \item \textbf{MRxD, a usage-pattern framework derived from a systematic analysis of 238 MR cross-device papers}, which characterizes systems through User Attention, Task Relatedness, and Cognitive Symmetry.
  \item \textbf{Six recurring usage patterns and associated design strategies} that explain how immersive and non-immersive devices are coordinated across different forms of MR cross-device activity.
  \item \textbf{An exploratory design workshop ($n=6$) and resulting design guidance}, providing initial evidence of how MRxD can support pattern-based design exploration and help designers identify overlooked design opportunities.
\end{enumerate}

\section{Background}

\subsection{Cross-Device Interaction}

Cross-device interaction broadly concerns interactions that extend beyond the boundary of an individual device by coordinating multiple devices, displays, or interaction resources within a shared activity or device ecology~\cite{cd-taxonomy}. Research in this area has developed over several decades, from multi-monitor workspaces~\cite{cd-multi-screen} and augmented display environments~\cite{rekimoto1999augmented} to increasingly mobile and ad-hoc device ecologies~\cite{radle2014huddlelamp,mrcd028,mrcd038}. This long development has also produced several overlapping research traditions. For example, \emph{distributed user interfaces} consider how interface components and interaction resources can be distributed across multiple computational and physical resources~\cite{demeure2005reference}, while multi-display research has examined how visual attention shifts across displays~\cite{rashid2012factors}. Related work has also been framed using overlapping terms such as
\emph{cross-surface interaction}~\cite{homaeian2017grounding},
\emph{multi-device interaction}~\cite{aumi2013doplink},
and \emph{cross-display interaction}~\cite{chung2018savil}. Although these traditions overlap and findings from one can often inform another, they developed with different scopes and terminology, leaving the broader research landscape conceptually fragmented. 
Brudy et al.~\cite{cd-taxonomy} synthesized these strands under the umbrella of \emph{cross-device computing}, providing a common vocabulary for interactions spanning diverse device configurations and ecologies. Their taxonomy organizes the cross-device design space along six dimensions, including \emph{temporal}, \emph{configuration}, \emph{relationship}, \emph{scale}, \emph{dynamics}, and \emph{space}, which collectively characterize how cross-device ecologies are organized, used, and changed over time. Because these dimensions are not tied to a particular device class, the taxonomy encompasses conventional displays, mobile and wearable devices, tangibles, and head-worn AR/VR devices.

Brudy et al.'s broad framing of cross-device interaction encompasses both homogeneous and heterogeneous combinations of devices. Within this broader landscape, HUIs, originally introduced by Feiner and Shamash~\cite{feiner-hui} as combinations of heterogeneous display and interaction device technologies, emphasize the complementary use of heterogeneous components. Focusing on the prevalent combination of conventional 2D components with MR environments, Hubenschmid et al.~\cite{mr-hui} characterize this subset as MR-HUIs. Their taxonomy builds on Brudy et al.'s cross-device framework, adapting its six dimensions and introducing \emph{interaction dynamics} and \emph{anchoring} to account for interaction and spatial relationships across MR and 2D components. Collectively, these taxonomies provide rich vocabularies for describing how cross-device systems are composed and organized across device ecologies.

\subsection{Usage-Oriented Perspectives in Cross-Device Interaction}

Beyond characterizing device configurations and interaction techniques, cross-device research has also examined how activities unfold across devices in use. Cross-device taxonomy has captured usage-facing characteristics such as temporal organization and user--device relationships~\cite{cd-taxonomy}, while empirical studies have examined task organization and attention across devices. For example, Jokela et al. distinguished sequential and parallel use as well as related and unrelated tasks~\cite{jokela2015diary}, while Rashid et al.~\cite{rashid2012factors} showed that display arrangement and content coordination can shape attention switching. In cross-reality settings, a prior work found that transitions among desktop, AR, and VR can introduce friction when they disrupt users' spatial mental models, highlighting the importance of maintaining spatial continuity across realities~\cite{mrcd003}.

To capture such behavior more systematically, prior work has also organized recurring cross-device activities into usage patterns. Yuan et al.~\cite{irene-usage} studied everyday multi-device work practices and identified five recurring patterns: \emph{integrating}, \emph{cloning}, \emph{expanding}, \emph{partitioning}, and \emph{migrating}. Their analysis positions usage patterns as a frame of reference for grounding cross-device technologies in workflows observed in practice. They further related these patterns to physical configurations, user motivations, and opportunities for cross-device design. Li et al.~\cite{cd-education} applied a similar usage-pattern perspective to in-the-wild cross-device learning while additionally examining attention allocation and shifting across devices. Their findings show how recurring workflows and attention issues can reveal design opportunities for cross-device learning platforms.

Together, these studies demonstrate the value of examining cross-device systems through how users coordinate tasks, attention, and transitions in use, complementing descriptions of how devices and interfaces are configured. However, existing usage-pattern studies have primarily focused on conventional non-immersive device ecologies. This motivates examining how such a usage-oriented perspective extends to interactions that combine immersive and non-immersive devices.

\subsection{Scope and Research Positioning}

For the purposes of this review, we use \textit{MR cross-device system} to refer to an HMD-based MR system used together with at least one physically separate or peripheral non-immersive device during a concrete user activity. 
We use \textit{MR cross-device} as a descriptive review term grounded in two lines of prior work: cross-device taxonomy, which studies interaction across device ecologies~\cite{cd-taxonomy}, and MR-HUI review, which focuses on complementary combinations of conventional 2D devices and MR environments~\cite{mr-hui}. 
Our review focuses on widely available, general-purpose non-immersive devices and peripherals, including phones, tablets, laptops, desktops, keyboards, mice, pens, and smartwatches. These devices recur across MR systems and often take clear roles in display, input, control, text entry, pointing, and sensing. This boundary helps us identify usage patterns and design strategies around general forms of cross-device coordination, while excluding highly specialized or one-off hardware whose interaction is mainly shaped by device-specific mechanisms.

Our scope differs from the MR-HUI review~\cite{mr-hui} in its organizing perspective and inclusion criteria. While that review primarily characterizes MR-HUI systems through device and component configurations, our review is organized around \textit{usage patterns}: how users coordinate immersive and non-immersive devices during concrete activities. Because our framework is intended to characterize recurring forms of cross-device coordination rather than particular MR-HUI component configurations, we use the broader term \textit{MR cross-device}. MR-HUI nevertheless provides an important basis for defining the systems examined in this review. We also retain physically distinct non-immersive devices when they shape the user's cross-device activity, even if they provide only input, output, or sensing support, as well as asymmetric collaborative setups in which users participate through different device arrangements within the same shared activity.

\section{Methods} \label{sec:method}

We conducted a PRISMA-informed systematic review~\cite{page2021prisma} to answer our RQs. %
Based on the identified papers, we developed a usage-centric framework through combined deductive and inductive process, and analyzed the design strategies associated with each usage pattern.

\begin{figure}
    \centering
    \includegraphics[width=0.95\linewidth]{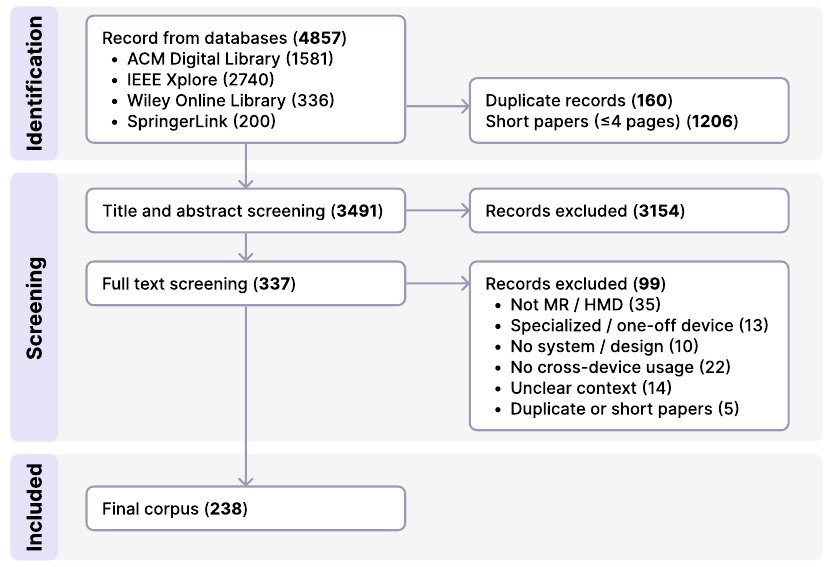}
    \caption{PRISMA diagram of paper collection and screening}
    \Description{
    A PRISMA flow diagram showing paper identification, screening, and inclusion. A total of 4,857 records were collected from ACM Digital Library (1,581), IEEE Xplore (2,740), Wiley Online Library (336), and SpringerLink (200). After removing 160 duplicates and 1,206 short papers, 3,491 records underwent title and abstract screening, of which 3,154 were excluded. The remaining 337 papers underwent full-text screening, with 99 excluded for not involving MR/HMD, non-daily devices, lack of a system or design, lack of cross-device usage, unclear context, or duplication/short length. The final corpus contained 238 papers.
    }
    \label{fig:prisma_flow}
\end{figure}

\subsection{Paper Collection}

We searched four digital libraries that publish substantial portions of HCI and MR research: ACM Digital Library, IEEE Xplore, Wiley Online Library, and SpringerLink. The search covered work published between January 2015 and December 2025 to capture recent MR cross-device systems and contemporary devices. We searched titles, abstracts, and keywords where supported by each digital library. For SpringerLink, whose advanced search does not support abstract-specific searches, we searched titles and keywords only. 
The search query comprised three keyword sets: a main set, a supplementary set, and a reality set. In abstract form, the query combined the main and supplementary sets with the reality set, while a small subset of terms in the main set that inherently describe hybrid or cross-reality interfaces were searched independently. 
The main set was informed by prior reviews of MR-HUI and cross-device taxonomies~\cite{mr-hui,cd-taxonomy}. It included terms directly related to MR cross-device interaction, such as ``cross-device,'' ``hybrid user interfaces,'' and ``multi device.'' Considering the fragmented terminology in this area~\cite{mr-hui,cd-taxonomy}, the supplementary set included device and interaction terms that could imply cross-device usage, such as ``phone,'' ``tablet,'' ``laptop,'' ``smartwatch,'' ``keyboard,'' ``mouse,'' ``pen,'' and ``asymmetric interaction.'' The reality set included ``augmented reality,'' ``virtual reality,'' ``mixed reality,'' and ``extended reality.'' The full search query is provided in Appendix~\ref{app:search-query}. The search returned 4,857 records\footnote{The search date was April 28th, 2026.}. We removed 160 duplicates and excluded papers of four pages or fewer. Because our analysis required sufficient descriptions of user activities and system design to analyze cross-device usage behavior, we used paper length as a pragmatic minimum-reporting threshold, following similar page-length criteria adopted in prior systematic reviews~\cite{ads-survey,xavier2024pseudo}. This process resulted in 3491 records for title and abstract screening.

\subsection{Screening and Eligibility Check}

Screening proceeded in two stages. First, we screened titles and abstracts to remove papers that were not written in English or were clearly outside the scope of the review. Specifically, we excluded papers that were (1) content-level duplicates (e.g., poster~\cite{mrcd193} of a full paper~\cite{mrcd136}), (2) unrelated to MR or MR head-mounted displays (e.g., handheld AR~\cite{mrcd293}), (3) centered on highly specialized or one-off devices whose roles were not transferable to common cross-device settings (e.g., haptic gloves~\cite{exclude-gloves}), or (4) did not propose a system or design for MR cross-device interaction (e.g., surveys~\cite{exclude-survey-1}). This stage left 337 papers for the second stage: full-text eligibility assessment. In addition to the criteria above, we excluded papers when the full text did not describe actual cross-device usage or interaction, such as papers focused only on backend algorithms~\cite{exclude-backend,mrcd272}, mock devices~\cite{mrcd010}, or comparisons between devices~\cite{mrcd298}. We also excluded papers when the usage context was unclear (i.e. without a concrete application scenario~\cite{mrcd006, mrcd119}). This process resulted in a final corpus of 238 papers\footnote{The full list of papers is available at \url{https://mrxd-paper.pages.dev/}.}.
Figure~\ref{fig:prisma_flow} provides a detailed breakdown of  the numbers of papers at different review stages.

Before screening, the first four authors discussed the exclusion criteria to establish a shared interpretation of the review scope. In the first stage, records were divided among the four authors and screened individually. In the second stage, three authors reviewed the assigned papers. Across both stages, uncertain cases were marked and discussed until the authors reached an agreement.

\subsection{Framework Development and Corpus Analysis}

\begin{figure*}
    \centering
    \includegraphics[width=\linewidth]{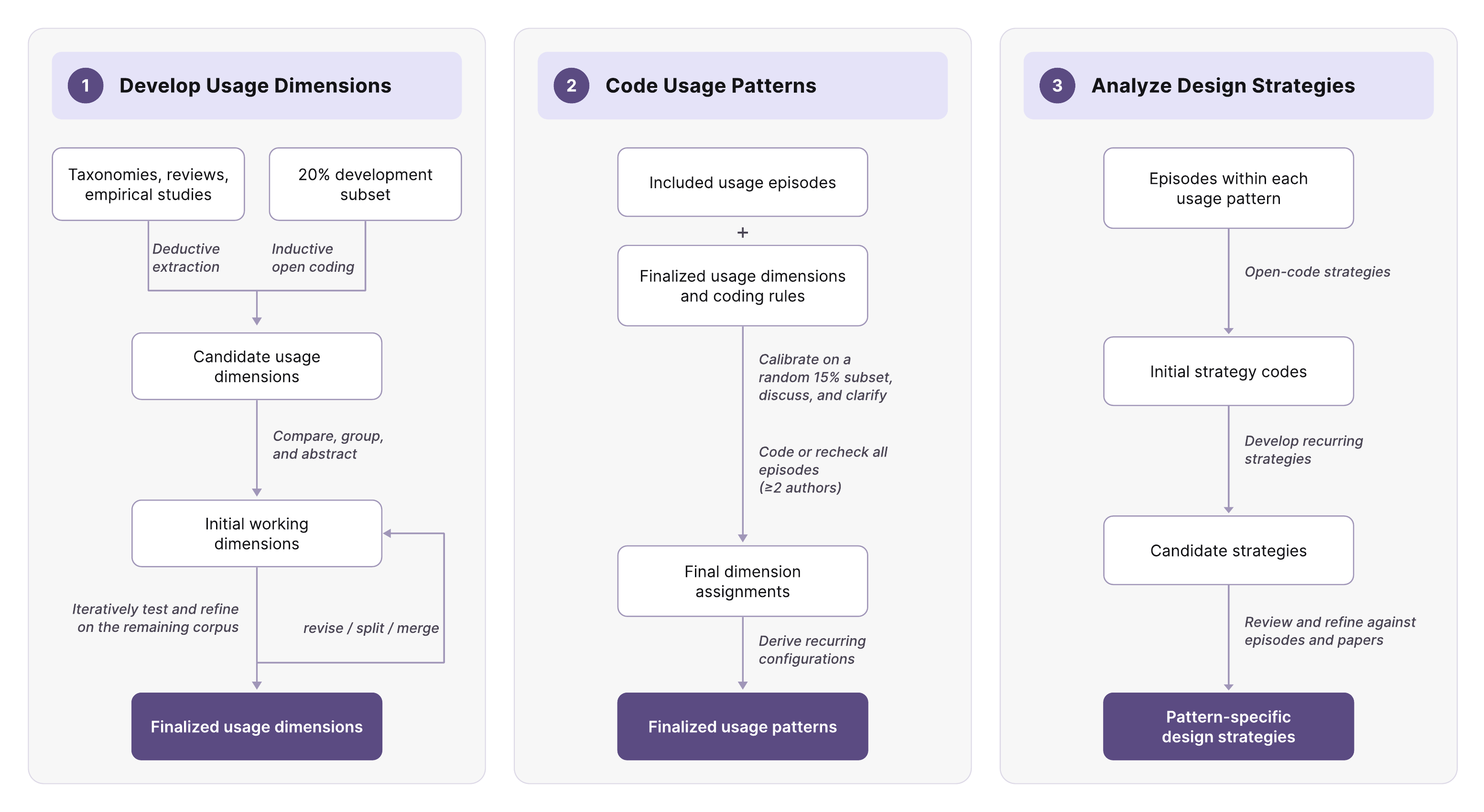}
    \caption{Overview of the MRxD framework development process. The analysis proceeded in three stages: (1) developing usage dimensions through deductive extraction from prior work and inductive open coding of a 20\% development subset, followed by iterative corpus-wide refinement; (2) coding included usage episodes with the finalized dimensions and coding rules, calibrating and clarifying before deriving recurring configurations as usage patterns; and (3) inductively analyzing episodes within each usage pattern, from initial strategy codes through recurring and refined strategies, to derive pattern-specific design strategies.}
    \Description{Flowchart showing three stages of the MRxD framework development process. 
    In Develop Usage Dimensions, taxonomies, reviews, and empirical studies inform deductive extraction, while a 20\% development subset is inductively open-coded. These sources produce candidate usage dimensions, which are compared, grouped, and abstracted into initial working dimensions. The dimensions are then iteratively tested on the remaining corpus and revised, split, or merged to produce finalized usage dimensions. 
    In Code Usage Patterns, included usage episodes are analyzed using the finalized usage dimensions and coding rules. The authors calibrate on a random 15\% subset, discuss and clarify disagreements, and then code or recheck all episodes with at least two authors. Final dimension assignments are used to derive recurring configurations, producing the finalized usage patterns. 
    In Analyze Design Strategies, episodes within each usage pattern are open-coded to produce initial strategy codes. Recurring strategies are then developed into candidate strategies, which are reviewed and refined against the episodes and papers to produce pattern-specific design strategies.}
    \label{fig:dev-process}
\end{figure*}

Our framework development and strategy synthesis were qualitative and iterative, supplemented by descriptive frequency and co-occurrence analyses. We first developed usage dimensions and used their configurations to identify recurring usage patterns (\textbf{RQ1}); we then analyzed how prior systems realized each pattern through design strategies (\textbf{RQ2}). Rather than coding usage patterns directly, we derived them from a small set of \emph{usage dimensions} (observable aspects along which MR cross-device use differs), where a \emph{usage pattern} is a recurring configuration of values across these dimensions. Coding dimensions first avoided grouping systems by surface similarities such as shared devices or applications, and instead made the criteria for distinguishing patterns explicit and consistent across a heterogeneous corpus. We analyzed patterns before strategies because a design strategy is a way of realizing a particular usage pattern; interpreting strategies therefore depends on first establishing the patterns they serve. As summarized in Figure~\ref{fig:dev-process}, this proceeded in three stages: developing usage dimensions, coding usage patterns, and analyzing design strategies.

The unit of analysis was the \emph{usage episode}: a continuous segment of activity in which the user's goal, involved devices, and interaction workflow remained stable. Before identifying or applying the dimensions, we segmented the reported workflows into candidate episodes based on observable changes in these features. For example, \cite{mrcd107} contributed two episodes: a complementary sequential usage and an in-situ interaction. This episode-level unit allowed us to distinguish different forms of cross-device use within the same paper or system.
We retained episodes central to a paper's design motivation or system contribution: those explicitly motivated, implemented, evaluated as a study condition, or discussed as a contribution. We excluded episodes that appeared only as comparative baselines~\cite{mrcd206} or as conventional interactions retained for system operation, such as standard typing beneath an augmented keyboard mapping~\cite{mrcd096}. When a paper gave equal attention to multiple techniques or scenarios, as in comparative studies~\cite{mrcd140} and design-space explorations~\cite{mrcd025}, we included all corresponding episodes. These episodes served as the cases for developing the MRxD framework. 

\subsubsection{Identifying Usage Patterns (RQ1)}

\paragraph{Developing usage dimensions.}
We developed the usage dimensions by alternating between conceptual (top-down) and empirical (bottom-up) analysis, informed by the iterative logic of taxonomy-development methods~\cite{nickerson2013}. Following this approach, we anchored the process in a \emph{meta-characteristic}, which provides the overarching basis for selecting relevant dimensions~\cite{nickerson2013}. Given our usage-centric goal, we defined this meta-characteristic as how users coordinate immersive and non-immersive devices within a usage episode. A candidate dimension was retained only if it (1) was a logical consequence of this meta-characteristic, that is, captured a user-facing difference in cross-device coordination rather than an implementation detail, (2) could be coded from the interaction workflows reported for an episode, and (3) distinguished meaningful differences across episodes that were not already captured by another dimension. Notably, this scoping is deliberate: the dimensions are not intended to describe every way MR cross-device systems differ, but to provide a parsimonious and explanatory lens tied to our analytical purpose. Consistent with the philosophy of iterative taxonomy development, our goal was a useful rather than a uniquely optimal set of dimensions~\cite{nickerson2013}; aspects that fell outside the meta-characteristic were excluded or kept as lower-level descriptors rather than elevated to dimensions.

We first assembled a broad set of candidate dimensions. Deductively, we drew usage-related dimensions from prior reviews, taxonomies, and empirical studies of cross-device use~\cite{mr-hui,irene-usage,cd-taxonomy,cd-education,ads-survey,cr-review}, such as attention, task properties, usage flow, interaction and information distribution, spatial and temporal relations, and device roles. Inductively, three authors independently open-coded the usage episodes identified in a random 20\% subset of the corpus to surface aspects not covered by prior work. We then compared, grouped, and abstracted these candidates into a small set of \emph{initial} working dimensions. For example, data similarity and interaction mapping first appeared as separate aspects, but both reflected how users must interpret the relationship between devices and were subsumed under a single provisional dimension.

One author then led an exploratory refinement pass over the remaining corpus to test whether the provisional dimensions and their distinctions could characterize usage episodes clearly and consistently. Dimension assignments made during this pass were functioned as diagnostic probes and used only to guide refinement. The other two authors reviewed the proposed revisions and flagged episodes, and all three authors discussed whether these cases required revised definitions, split or merged dimensions, or lower-level descriptors. Whenever a dimension changed, the affected episodes were revisited to test the revised formulation. For instance, we narrowed User Attention to overt visual attention, which could be inferred more consistently from the interaction behavior reported in papers. Drawing on Nickerson et al.'s use of explicit ending conditions~\cite{nickerson2013}, we finalized the dimensions when their definitions and distinctions remained stable through the final pass, all included episodes could be characterized, and no recurring residual cases required an additional independent dimension.

This produced three final usage dimensions, which we define and justify in Section~\ref{subsec:dimensions}: \emph{User Attention}, \emph{Task Relatedness}, and \emph{Cognitive Symmetry}. We operationalized each dimension as binary with two categories to support consistent coding across the corpus. A usage pattern is then defined by a recurring configuration of values across these dimensions. This binary treatment involved a trade-off: while it made large-corpus coding feasible, it simplified experiences that may vary along a spectrum. We therefore treat the dimensions as design-oriented analytical categories for characterizing recurring MR cross-device use rather than as a closed taxonomy, and return to this limitation in Section~\ref{sec:discussion}.

\paragraph{Coding usage patterns.}\label{subsubsec:pattern-coding}
After finalizing the dimensions and their operational definitions, we conducted formal pattern coding across the corpus. Unlike the exploratory assignments used during dimension refinement, this stage aimed to produce the final and consistent pattern assignments using a fixed set of definitions and decision rules. Earlier assignments made during dimension refinement were not treated as final results; each episode was coded or rechecked against the finalized codebook. 
For each included episode, the three author assigned the applicable dimension values and derived the corresponding pattern. 
To calibrate the finalized coding scheme, the three authors independently coded the usage episodes from a random 15\% subset papers. Pairwise Cohen's $\kappa$ exceeded .70 across all three author pairs, indicating substantial agreement~\cite{mchugh2012interrater}. After discussing disagreements and clarifying the rules for edge cases, all other episodes were coded or rechecked against the final codebook by at least two authors, with disagreements resolved through discussion. 
For corpus-level prevalence and co-occurrence analyses, repeated instances of the same pattern within a paper were counted once. Distinct implementations were retained as separate cases for design-strategy analysis.

\subsubsection{Analyzing Design Strategies (RQ2)}

To understand how each pattern was instantiated in prior systems, we analyzed the design approaches within each pattern using reflexive thematic analysis~\cite{Braun01012006}, following prior review practices for synthesizing design strategies~\cite{luo2026beyond,alvarado2022systematic}. The unit of analysis remained the usage episode, and an episode could contribute multiple strategies. One author open-coded the strategies across all usage episode and merged similar designs into higher-level categories; the other authors reviewed these categories against the associated papers to ensure that each reflected recurring design choices.

Alongside coding, we extracted structured metadata describing system context, device and user configurations, cross-device interaction and coordination, and physical arrangements, partly adapted from prior cross-device taxonomies~\cite{mr-hui,cd-taxonomy} and from the lower-level descriptors identified during dimension development. We treated these as descriptive context for the usage patterns rather than as the primary basis of the framework. Descriptive distributions of these metadata are provided in the Supplementary Material. For co-occurrence analyses, we used lift to normalize observed frequencies by category prevalence. This distinguished associations driven by the prevalence of common patterns or devices from those occurring more or less often than expected under independence. We applied lift to both pairwise pattern combinations and associations between device type and usage pattern.

\section{MRxD Framework}

\begin{figure*}
    \centering
    \includegraphics[width=\linewidth]{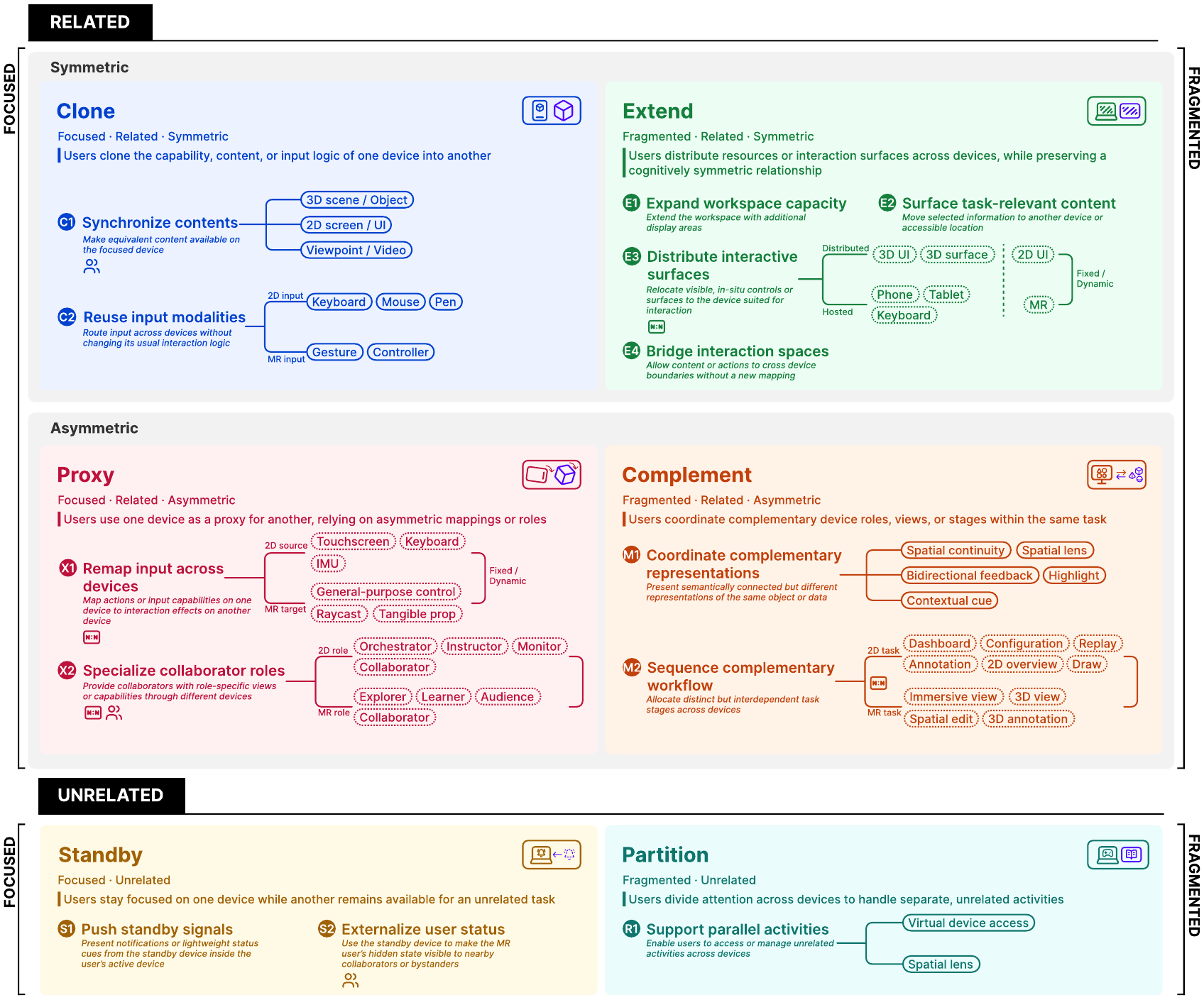}
    \caption{MRxD framework. The horizontal axis represents User Attention (Focused to Fragmented). The framework first separates Task Relatedness into Related and Unrelated use. Within Related use, patterns are further distinguished by Cognitive Symmetry (Symmetric vs. Asymmetric). Because Cognitive Symmetry is not applicable to Unrelated tasks, the framework yields six usage patterns rather than eight combinations. \iconUser{} indicates collaborative cases, and \iconMtm{} indicates many-to-many mapping from two branches. The purple icons represent the virtual objects in MR.}
    \Description{
    A framework diagram with six usage-pattern panels. The Related section forms a two-by-two layout: Clone at the upper left, Extend at the upper right, Proxy at the lower left, and Complement at the lower right. Clone includes strategies for synchronizing contents and reusing input modalities; Extend includes expanding workspace capacity, surfacing task-relevant content, distributing interactive surfaces, and bridging interaction spaces; Proxy includes remapping input across devices and specializing collaborator roles; and Complement includes coordinating complementary representations and sequencing complementary workflows. The Unrelated section contains Standby on the left, with strategies for pushing standby signals and externalizing user status, and Partition on the right, which supports parallel activities. Branch diagrams within the panels illustrate representative mappings between devices, MR content, views, roles, and task stages.
    }
    \label{fig:mrxd-main}
\end{figure*}

\begin{figure*}
    \centering
    \includegraphics[width=0.9\linewidth]{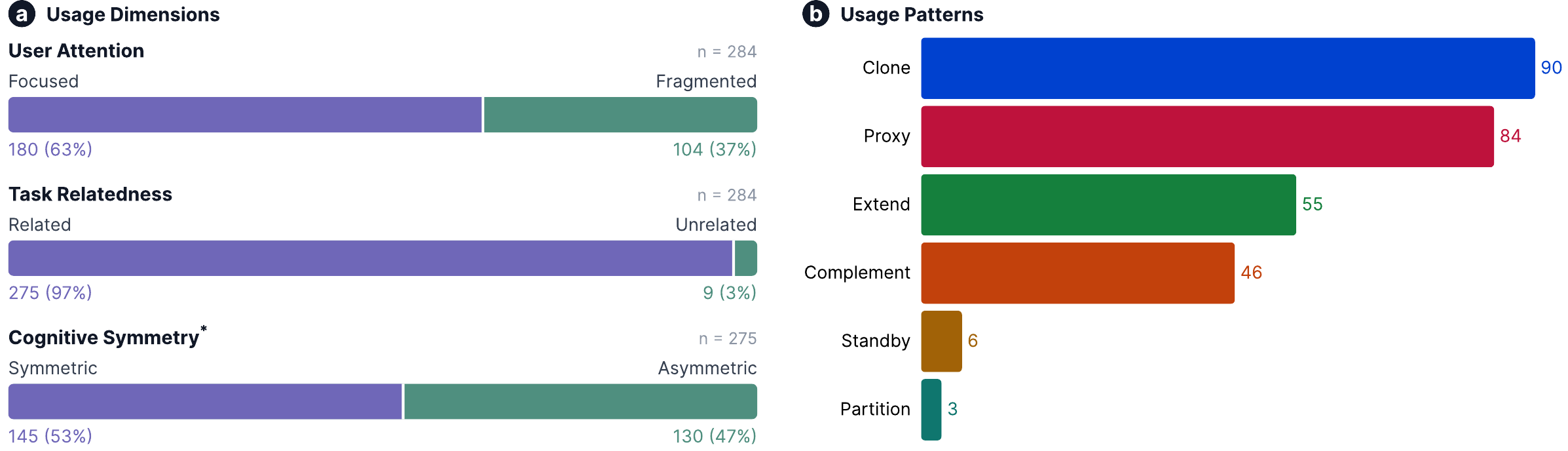}
    \caption{Prevalence of usage dimensions and usage patterns in the corpus. (a) Distribution of usage episodes across the three usage dimensions. Attention and Task Relatedness are reported for all 284 episodes, while Cognitive Symmetry is defined only for the 275 Related episodes. (b) Number of usage episodes exhibiting each usage pattern.}
    \Description{Two-panel figure summarizing the prevalence of usage dimensions and usage patterns. Panel (a) contains three horizontal proportion bars. For Attention, 180 of 284 usage episodes (63\%) are Focused and 104 (37\%) are Fragmented. For Task Relatedness, 275 of 284 episodes (97\%) are Related and 9 (3\%) are Unrelated. Cognitive Symmetry is evaluated only for the 275 Related episodes, of which 145 (53\%) are Symmetric and 130 (47\%) are Asymmetric. Panel (b) is a horizontal bar chart showing the number of usage episodes associated with each usage pattern: Clone, 90; Proxy, 84; Extend, 55; Complement, 46; Standby, 6; and Partition, 3.}
    \label{fig:prevalence}
\end{figure*}

Through the process described in Section~\ref{sec:method}, we propose \textbf{MRxD}, a usage-pattern framework for analyzing MR cross-device systems from a user-centered perspective. As shown in Figure~\ref{fig:mrxd-main}, MRxD is built around three usage dimensions---\textit{User Attention}, \textit{Task Relatedness}, and \textit{Cognitive Symmetry}. Different configurations of these dimensions define recurring usage patterns. For task-related episodes, crossing User Attention and Cognitive Symmetry yields four configurations: Clone, Proxy, Extend, and Complement. For task-unrelated episodes, Cognitive Symmetry is not applicable because the devices do not support a shared task, leaving two configurations distinguished by User Attention: Standby and Partition. For each pattern, we further identify design strategies that explain how prior systems supported that pattern in practice.

Figure~\ref{fig:prevalence} provides an overview of how these dimensions and usage patterns were distributed across the reviewed corpus. Most usage episodes involved task-related device use, while attention and cognitive symmetry showed more balanced distributions. At the pattern level, Clone and Proxy were the most prevalent, followed by Extend and Complement, whereas Standby and Partition occurred only occasionally.

\subsection{Usage Dimensions} \label{subsec:dimensions}

\subsubsection{User Attention} \label{subsubsec:attention}

User Attention describes whether users distribute visual attention across immersive and non-immersive devices during a usage episode. Prior work treats visual attention as a key design trade-off: additional displays can expand available resources, but may also introduce attention-switching costs or distraction~\cite{rashid2012factors,cd-education}. Attention distribution also helps reveal which device acts as the activity foreground~\cite{mrcd011} and where designers may need to reduce or guide context switching~\cite{cd-education}. In MRxD, we focus on \emph{overt} visual attention, observable through gaze or head orientation, rather than \emph{covert} attention through peripheral perception~\cite{theory-attention}, because overt attention is more observable in system descriptions and easier to operationalize for analyzing cross-device coordination. \footnote{Unless otherwise stated, visual attention in this paper refers to overt visual attention.} %
We distinguish two forms: a usage episode is \textbf{Focused} when visual attention is directed to one device while another plays a supporting role, and \textbf{Fragmented} when users must visually attend to multiple devices to complete the task. The criterion is not switching frequency, but whether multiple spaces are visually necessary for task progress. 
For example, a sequential and exclusive workflow~\cite{mrcd124,mrcd191} is considered as fragmented since users must move visual attention between device spaces. Mirrored output~\cite{mrcd239} to MR is also treated as fragmented when the content remains tied to the non-immersive device and requires coordination with the MR view.

Across the reviewed systems, focused overt attention was enabled in two broad ways. In many cases, the secondary device remained part of the interaction, but users could operate it through \textbf{tactile or peripheral attention} instead of direct visual inspection. This was common when MR systems were paired with keyboards, mice, or pens, or touchscreens~\cite{mrcd022,mrcd023,mrcd086,mrcd137,mrcd240}, where users relied on touch~\cite{mrcd025}, motor memory~\cite{mrcd094}, or simple repeated gestures~\cite{mrcd122,mrcd118}. 
In other cases, the cross-device configuration required \textbf{attention to only one device}. This occurred when relevant resources were duplicated, mirrored, or placed directly in the user's current view, removing the need to inspect the source device~\cite{mrcd023,mrcd013}; when another device operated in the background by supplying sensed information without requiring user interaction~\cite{mrcd223}; or in collaborative settings where responsibilities were distributed across users, allowing each person to remain focused on their own device~\cite{mrcd048,mrcd012,mrcd136}.

Two main reasons appeared for fragmented usage. For \textbf{resource access}, attention shifted because information needed for the ongoing task was distributed across devices. A common case involved moving between 2D screen-based content and 3D spatial content to understand correspondences~\cite{mrcd107,mrcd077,mrcd004}, either across different representations of the same data~\cite{mrcd077,mrcd305} or across different data entities within one activity~\cite{mrcd169}. Fragmented attention also occurred in multi-screen settings, where users shifted between one physical and multiple virtual 2D displays to access distributed resources~\cite{mrcd285}. Resource access could involve peeking behaviors, where users briefly returned to another device to compare information or confirm its status~\cite{mrcd003,mrcd017}. Data transfer could likewise fragment attention when users accessed and moved a digital object across devices~\cite{mrcd019,mrcd196}.
For \textbf{control access}, attention shifted in two main situations. First, some control interfaces themselves required visual engagement, and thus users needed to look at the device to guide actions such as typing on touchscreen keyboards~\cite{mrcd024}, sketching precisely~\cite{mrcd037}, adjusting parameters~\cite{mrcd003,mrcd107}, or manipulating on-screen controls~\cite{mrcd026,mrcd316}. Second, some interactions created a cross-device dependency between input and output, requiring users to act on one device and then monitor or assess the corresponding result on another. In some systems, such switching was deliberately built into the cross-device interaction~\cite{mrcd024,mrcd026,mrcd077}, and could also become embedded in users' own workflows as they coordinated actions and inspection across devices~\cite{mrcd017}.

\subsubsection{Task Relatedness}

Task Relatedness describes whether tasks on immersive and non-immersive devices are connected. Tasks are \textbf{Related} when the devices contribute to a single overarching goal, even if each device supports a different subtask, representation, or interaction role. Tasks are \textbf{Unrelated} when the devices support different goals or parallel activities. This dimension extends prior non-MR cross-device work~\cite{irene-usage} and helps explain when devices are used together for one activity and when MR cross-device creates multitasking or coordination challenges across separate activities.

When tasks were related, we identified several recurring device relationships. In an \textbf{input medium} relationship, one device supplied input for the same task on another, through either user actions or sensed information. User-operated input flowed from non-immersive to immersive devices~\cite{mrcd045,mrcd240}, or in the reverse direction~\cite{mrcd025,mrcd031}, through modalities such as touch~\cite{mrcd045}, typing~\cite{mrcd038}, drawing~\cite{mrcd049}, mouse movement~\cite{mrcd101}, and gesture~\cite{mrcd031}. Devices also supplied sensed information, including physiological signals~\cite{mrcd139,mrcd223}, voice~\cite{mrcd167}, and video streams~\cite{mrcd116,mrcd309}. 
In \textbf{collaboration} relationships, multiple users coordinated a shared task through different devices, often with asymmetric views~\cite{mrcd030}, differentiated roles~\cite{mrcd163}, or complementary interaction capabilities~\cite{mrcd136}. In \textbf{auxiliary display} relationships, one device presented information that users consulted to support task activities on another, such as references~\cite{mrcd196,mrcd018}, previews~\cite{mrcd024,mrcd191}, or supplementary views~\cite{mrcd194,mrcd200}. \textbf{Workflow chain} relationships captured cases where users advanced the same task across stages, moving materials, decisions, or actions from one device to another sequentially~\cite{mrcd003,mrcd197,mrcd124,mrcd318}. %

Unrelated tasks appeared less frequently, with two main forms. In \textbf{multitasking}, users maintained separate activities across immersive and non-immersive devices, such as messaging or watching video independently of the ongoing immersive task~\cite{mrcd154,mrcd044}. In \textbf{standby} relationships, one device remained available as a background channel for notifications or an outward-facing channel for communication and social awareness, while the primary user stayed engaged in an immersive activity~\cite{mrcd011,mrcd128,mrcd095,mrcd159}.

\subsubsection{Cognitive Symmetry} \label{subsubsec:cogsym}

Cognitive Symmetry describes whether the user can carry the same interpretation across the devices, or must construct a new interpretive mapping to relate a representation or action on one device to its counterpart on another.
A relationship is \textbf{Symmetric} when the correspondence is semantically direct, so users reuse an existing interpretation across devices without an additional interpretive mapping. It is \textbf{Asymmetric} when they must reinterpret how a representation or action on one device corresponds to information or effects on another.
This distinction of Cognitive Symmetry is informed by the notion of semantic distance, which characterizes how directly interface expressions correspond to users' intentions and the interpretive transformation required to bridge them~\cite{theory-gulf}. We extend this underlying intuition from user--interface mappings to cross-device correspondences involving representations and actions. 
Cognitive Symmetry matters because immersive systems often combine 3D spatial representations with conventional 2D interfaces and devices, which can create gaps in how information is represented, aligned, and manipulated across devices. Prior work has suggested that spatial mental models can help users predict outcomes and reduce friction across reality~\cite{mrcd003}. Cognitive Symmetry helps explain when designers should preserve direct cross-device correspondence and when clearer metaphors, alignment cues, or device-specific interaction designs may be needed.

To operationalize this distinction in our coding, we considered both representational and action--effect relationships. For representations, a change in device or display location does not by itself introduce asymmetry when the underlying representational structure is preserved; for example, a conventional 2D interface extended from a desktop into an MR virtual screen can still be interpreted in the same way. In contrast, when information is recast into a different representational structure, such as from a 2D visualization to a spatial 3D representation, users must interpret how elements and relations in one form correspond to those in the other. The same principle applies to action--effect relationships: asymmetry arises when an action on one device must be reinterpreted to understand its effect on another, rather than carrying over through an already familiar correspondence.

In the reviewed systems, symmetric relationships were often relatively straightforward. For representations, devices preserved similar representational forms, such as physical and virtual screens~\cite{mrcd285,mrcd234}, or presented the same or semantically equivalent content, including shared documents~\cite{mrcd023,mrcd200}, snapshots~\cite{mrcd001,mrcd196}, and views~\cite{mrcd005,mrcd174}. For actions, familiar inputs retained their conventional effects, such as typing text with a keyboard~\cite{mrcd209}, drawing strokes with a pen~\cite{mrcd049}, or selecting a button by tapping it~\cite{mrcd149}.
Asymmetric relationships instead introduced a new mapping between structures across devices. For representations, information could be organized or expressed in different forms. This commonly occurred when 3D spatial content was paired with 2D parameters, lists, charts, or abstract views~\cite{mrcd003,mrcd017,mrcd107,mrcd124,mrcd305}. For actions, users needed to interpret how an action on one device was mapped to an effect elsewhere, including what it controlled and where the resulting change appeared. For example, a smartphone could act as an MR controller~\cite{mrcd025,mrcd194}, a trackpad for a virtual cursor~\cite{mrcd094}, or a spatial proxy for an object~\cite{mrcd114,mrcd154}.

Across these asymmetric cases, we observed two broad ways in which users could bridge the resulting correspondence, including \textbf{explicit bridging} and \textbf{implicit bridging}. These mechanisms are not mutually exclusive; users may combine several of them during actual use. 
Explicit bridging relied on external support that made the cross-device relationship more observable or interpretable. One form is \textbf{system alignment}, where the interface provides perceptual cues that expose the correspondence between representations or action effects (e.g. bidirectional highlighting)~\cite{mrcd035,mrcd028,mrcd255}. When such cues are absent or insufficient, users may need \textbf{instructional scaffolding}, such as tutorials, onboarding, system prompts, researcher guidance~\cite{mrcd166,mrcd112,mrcd122,mrcd107}. Under multi-user settings, users may also rely on \textbf{collaborative mediation}, where verbal communication~\cite{mrcd004,mrcd136}, monitoring~\cite{mrcd163}, or shared annotations~\cite{mrcd111,mrcd184} help reconcile differences between device views and roles.

Implicit bridging occurs when users infer an otherwise asymmetric cross-device relationship without relying on explicit external support, instead drawing on prior knowledge or the immediate task context. One form relies on \textbf{established mappings}, where users apply a correspondence already familiar from prior experience. Although familiar, such mappings still require users to relate representations across different domains and are therefore coded as asymmetric. Examples include interpreting a 2D projection of a 3D object~\cite{mrcd037}, or connecting a 2D floor plan to a 3D space~\cite{mrcd184}. In other cases, users transfer a relational schema from a different familiar context to the cross-device setting. Following Structure-Mapping Theory (SMT)~\cite{theory-smt}, this \textbf{analogical transfer} relies on aligning relational roles across domains, such as treating a phone as a handheld controller~\cite{mrcd194} or a pen as a pointer~\cite{mrcd055}. Users may also rely on \textbf{contextual grounding} to infer the relationship. For example, when users revise parameters or code on a desktop and immediately inspect the resulting change in MR, the temporal and interaction context helps them infer how the desktop action relates to MR~\cite{mrcd124,mrcd169}.

\subsection{Usage Patterns and Design Strategies} \label{subsec:patterns}

\subsubsection{\textbf{\textcolor{clonecolor}{Clone}}: Focused, Related, and Symmetric}

In the Clone pattern $(n=90)$, users clone the capability, content, or input logic of one device into another while remaining focused on a primary device. The devices support the same task, and the cross-device relationship is cognitively symmetric because users can understand the relationship directly through the duplication. Among the usage patterns, Clone represents a foundational motivation for cross-device usage: it serves both as a basic synchronization mechanism and as a fallback or alternative interaction mechanism that retains existing device functions alongside more elaborate cross-device capabilities. However, this copy-and-paste logic can sometimes limit adaptation to the affordances of the receiving device. Two design strategies were used to realize the Clone pattern:

\begin{figure}[h]
    \centering
    \includegraphics[width=0.9\linewidth]{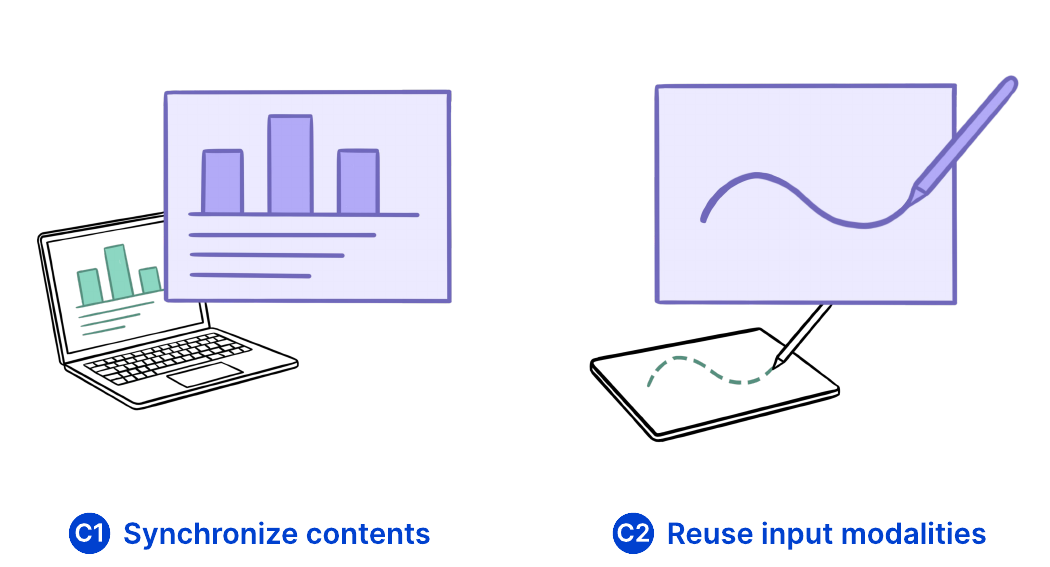}
    \caption{Illustrations of design strategies in the Clone pattern. 
    (C1) Synchronize contents: content displayed on a laptop is replicated in an MR window. 
    (C2) Reuse input modalities: pen input performed on a tablet is routed to a virtual canvas in MR.}
    \Description{Two side-by-side line drawings. In panel (C1), a laptop displaying a green bar chart sits behind a larger purple virtual window containing the corresponding chart. In panel (C2), a stylus is used on a tablet while a purple virtual canvas above it displays the resulting curved stroke.}
    \label{fig:illustration-clone}
\end{figure}

\designstrategy[clonecolor]{C1}{Synchronize content or views}
The system clones the same or equivalent content across devices so that users can build common ground, access resources without returning to the original device, or allow another user to observe the same activity. This includes shared 3D scenes for collaboration~\cite{mrcd012,mrcd008,mrcd016}, shared first-person views~\cite{mrcd048,mrcd103} or video streams~\cite{mrcd092,mrcd093} for embodied understanding, and shared 2D screens or UIs that are reproduced in MR~\cite{mrcd023,mrcd063,mrcd013}.  
In single user setting, designers can use this strategy when they want users to peek, adjust, or continue a 2D activity while keeping their visual attention within the MR workspace, such as by cloning a 2D screen or interface into MR~\cite{mrcd023,mrcd063}. In collaborative settings, cloned first-person views provide external users with a straightforward way to follow the MR user's view or egocentric perspective~\cite{mrcd048,mrcd092,mrcd103}. However, such views may provide limited surrounding context, which can be complemented by a shared 3D scene~\cite{mrcd103} and asymmetric information (see~\ref{subsubsec:proxy}).

\designstrategy[clonecolor]{C2}{Reuse input modalities}
This strategy retains the input function of one device while routing it to another device or environment without changing its interaction logic. Because these inputs can often be used without shifting visual attention, they help users remain focused on the primary display. For example, non-immersive input can be reused in immersive applications through keyboards~\cite{mrcd022,mrcd221}, mice~\cite{mrcd233,mrcd082}, or tablet-and-pen input~\cite{mrcd012,mrcd327}. This direction is useful for addressing limitations of current MR input, such as the lower efficiency or precision of controllers and hand gestures for certain tasks. Conversely, immersive input can be reused to operate non-immersive interfaces, such as using gestures~\cite{mrcd031,mrcd025} or controllers~\cite{mrcd013} to interact with a phone or desktop. In both directions, designers can further leverage this strategy to avoid unnecessary switches to a different input modality. This can be useful for brief, lightweight cross-device adjustments where switching input methods would add unnecessary overhead.

\subsubsection{\textbf{\textcolor{proxycolor}{Proxy}}: Focused, Related, and Asymmetric} \label{subsubsec:proxy}

In the Proxy pattern $(n=84)$, users remain visually focused on one device while another device supports the same task through an asymmetric input mapping or role relation. Unlike Clone, which preserves a shared representation or interaction logic across devices, Proxy assigns devices different but corresponding functions. For example, a handheld device may provide mapped control over spatial content in MR, while collaborative systems may present role-specific views of the same activity across devices. This distinction allows designers to tailor interaction, information, or user roles to the characteristics of each device. However, these differentiated functions require users to understand how their actions and representations correspond across devices. Such relationships can create cognitive gaps, making the bridging mechanisms described in Section~\ref{subsubsec:cogsym} particularly relevant.
\begin{figure}[h]
    \centering
    \includegraphics[width=0.9\linewidth]{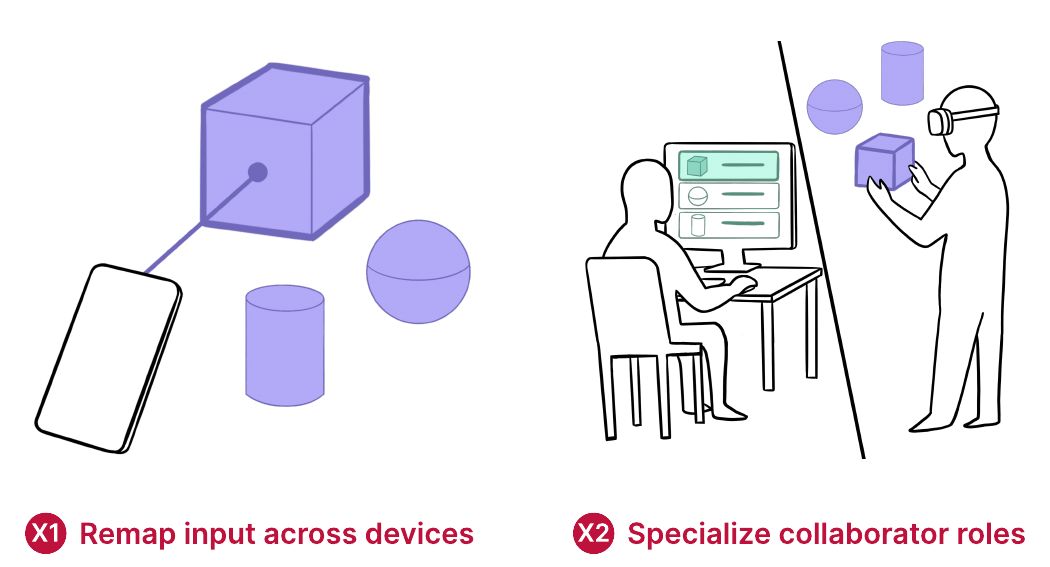}
    \caption{Illustrations of design strategies in the Proxy pattern.
    (X1) Remap input across devices: input from a smartphone is mapped to the manipulation of virtual objects in MR.
    (X2) Specialize collaborator roles: a desktop user guides an MR user through a virtual assembly task, with role-specific interfaces supporting task guidance on the desktop and direct manipulation in MR.}
    \Description{Two side-by-side line drawings. In panel (X1), a smartphone is connected to a purple virtual cube, illustrating smartphone input mapped to the manipulation of virtual objects. In panel (X2), two collaborators engage with the same virtual assembly task through different role-specific interfaces: a seated desktop user views task information and guidance on a monitor, while a standing MR user wearing a headset directly manipulates a purple virtual component.}
    \label{fig:illustration-proxy}
\end{figure}

\designstrategy[proxycolor]{X1}{Remap input across devices}
The physical affordances or input capabilities of one device are mapped to interaction effects on another. Similar to C2, interaction with the secondary device relies on proprioceptive, or peripheral cues, allowing users to keep their gaze on the immersive view. These mappings draw on IMU data~\cite{mrcd118}, touch input~\cite{mrcd107}, or keyboard actions~\cite{mrcd166}, and support interactions including 3D object manipulation~\cite{mrcd026,mrcd045}, raycasting~\cite{mrcd052,mrcd194}, tangible props~\cite{mrcd114,mrcd154}, and context-aware commands~\cite{mrcd240,mrcd299,mrcd122}.
Within this broad design space, input types and interaction effects could be combined flexibly. We further observed several recurring pairings. Continuous inputs based on IMU or touchscreen movement were often paired with continuous spatial operations such as object manipulation~\cite{mrcd118,mrcd240,mrcd045,mrcd011}. In contrast, discrete inputs, such as touchscreen gestures or touches at specific device regions, were more commonly mapped to invoking specific MR functions and shortcuts~\cite{mrcd107,mrcd122}.
When leveraging this strategy, one design consideration is how users understand the asymmetric mapping between input and effect. Designers can reduce this cognitive gap by reusing mappings established in everyday device interaction or gaming, allowing users to infer the new action--effect relationship through analogical transfer as described by SMT.
In our corpus, we found no clear cases in which MR input served as a proxy for interaction on a non-immersive device.

\designstrategy[proxycolor]{X2}{Specialize collaborator roles}
In collaborative settings, different devices support distinct roles while contributing to the same shared task. These systems align each role with the strengths of its device. Non-immersive devices can support information-dense or overview-oriented work through text, dashboards, timelines, and other 2D interfaces, making them suitable for roles such as orchestrators or instructors~\cite{mrcd030,mrcd088,mrcd136,mrcd111,mrcd163}. Immersive devices, in contrast, provide spatial, embodied, and situated access to the task environment, supporting roles such as learners or explorers~\cite{mrcd030,mrcd088,mrcd136,mrcd111,mrcd163}. This specialization allows collaborative systems to combine the output and interaction strengths of heterogeneous devices. However, users must understand not only their own role, but also what information and actions are available to others and their effects on task state. Designers should therefore expose the collaborator actions and task states needed for coordination while minimizing unnecessary monitoring and attention shifts~\cite{mrcd111,mrcd136}.

\subsubsection{\textbf{\textcolor{extendcolor}{Extend}}: Fragmented, Related, and Symmetric}

In the Extend pattern $(n=55)$, users distribute task-related resources, interfaces, or interaction surfaces across devices. The devices support the same activity, but users need to visually attend across them because each device provides task-relevant content or interaction opportunities, which distinguishes Extend from Clone. The cross-device relationship remains cognitively symmetric: users can understand the extension as additional space, equivalent content, direct and continuous controls.
Across the reviewed systems, Extend reflects two recurring directional motivations. When task resources are extended into MR, designers leverage the spatial flexibility of immersive space to overcome the fixed display area of physical screens, creating additional room for views, references, and controls. This flexibility, however, introduces a coordination challenge as resources must be arranged so that they remain easy to process and their relationships remain perceptually clear without making attention shifts across a broader workspace unnecessarily burdensome. In the reverse direction, designers extend UIs or interactive surfaces onto non-immersive devices to make use of stable, directly manipulable physical surfaces that can remain readily available alongside the immersive workspace.

\designstrategy[extendcolor]{E1}{Expand workspace capacity}
The system provides additional display space through user-anchored, device-anchored, or space-anchored virtual screens. This strategy also echos the design space of augmented displays~\cite{ads-survey}. The extended content may serve as independent resources~\cite{mrcd285,mrcd005,mrcd239}, or supplements to the main display~\cite{mrcd200,mrcd038,mrcd110}. When extending into MR, the immersive space serves as a large workspace for placing virtual screens, references, or resources around physical devices, resembling multi-display work~\cite{cd-multi-screen,rashid2012factors} but with greater flexibility in spatial placement, scale, and anchoring~\cite{mrcd285,mrcd239,mrcd234}. One work extended the workspace to the non-immersive device, using the 2D tablet to provide an extended 3D view of the MR virtual object (i.e. a fish tank view)~\cite{mrcd174}. Such design also benefits from the Proxy to leverage the tablet surface and pen for efficient selection. %

\designstrategy[extendcolor]{E2}{Surface task-relevant content}
This strategy is motivated by the need to reduce the effort of locating or monitoring information during an ongoing task. Systems support this by bringing frequently consulted resources, such as reference materials~\cite{mrcd196} or quick-access information~\cite{mrcd001}, into more accessible positions, for example by relocating them into virtual space or placing them near the main task area. Although this strategy facilitates information retrieval, it shifts part of the interaction cost to the selection and organization of relevant content, especially when information needs change frequently. Designers therefore need lightweight mechanisms for moving content across devices and organizing surfaced information without introducing excessive interaction overhead or spatial clutter~\cite{mrcd001,mrcd196}.

\designstrategy[extendcolor]{E3}{Distribute interactive surfaces}
The system relocates interactable surfaces to the device best suited for interaction. Since many MR controls remain familiar 2D UIs, such as menus~\cite{mrcd114,mrcd149}, tool palettes~\cite{mrcd120,mrcd194}, or shortcuts~\cite{mrcd206,mrcd096}, they can be redistributed across the device ecology: to tablets~\cite{mrcd014,mrcd120}, phones~\cite{mrcd168,mrcd194}, or keyboards~\cite{mrcd206,mrcd096}, whereas mobile controls can be reversely reproduced in MR as added-on interfaces~\cite{mrcd025,mrcd194}. Placing these interfaces on physical devices can exploit direct and precise input surfaces while keeping frequently used controls readily accessible. Such devices are often positioned within easy reach or integrated into the user's working posture, reducing the physical transition required to access them during immersive work.
This distribution may also be adaptive, with controls switching across surfaces according to context or user behavior, such as gaze~\cite{mrcd005}, proximity~\cite{mrcd149}, or interaction focus~\cite{mrcd168,mrcd194}. Beyond UI controls, task-specific surfaces, such as a 3D plane or object surface, may be relocated to a tablet to leverage precise 2D drawing~\cite{mrcd049,mrcd105,mrcd239}. Compared with Clone, these UIs usually contain visual elements that users need to check before acting, which leads to fragmented attention. Compared with Proxy, the interaction remains symmetric when the UI elements preserve direct action-effect relationships, such as tapping a visible button or drawing on a tablet surface.

\begin{figure}[h]
    \centering
    \includegraphics[width=0.9\linewidth]{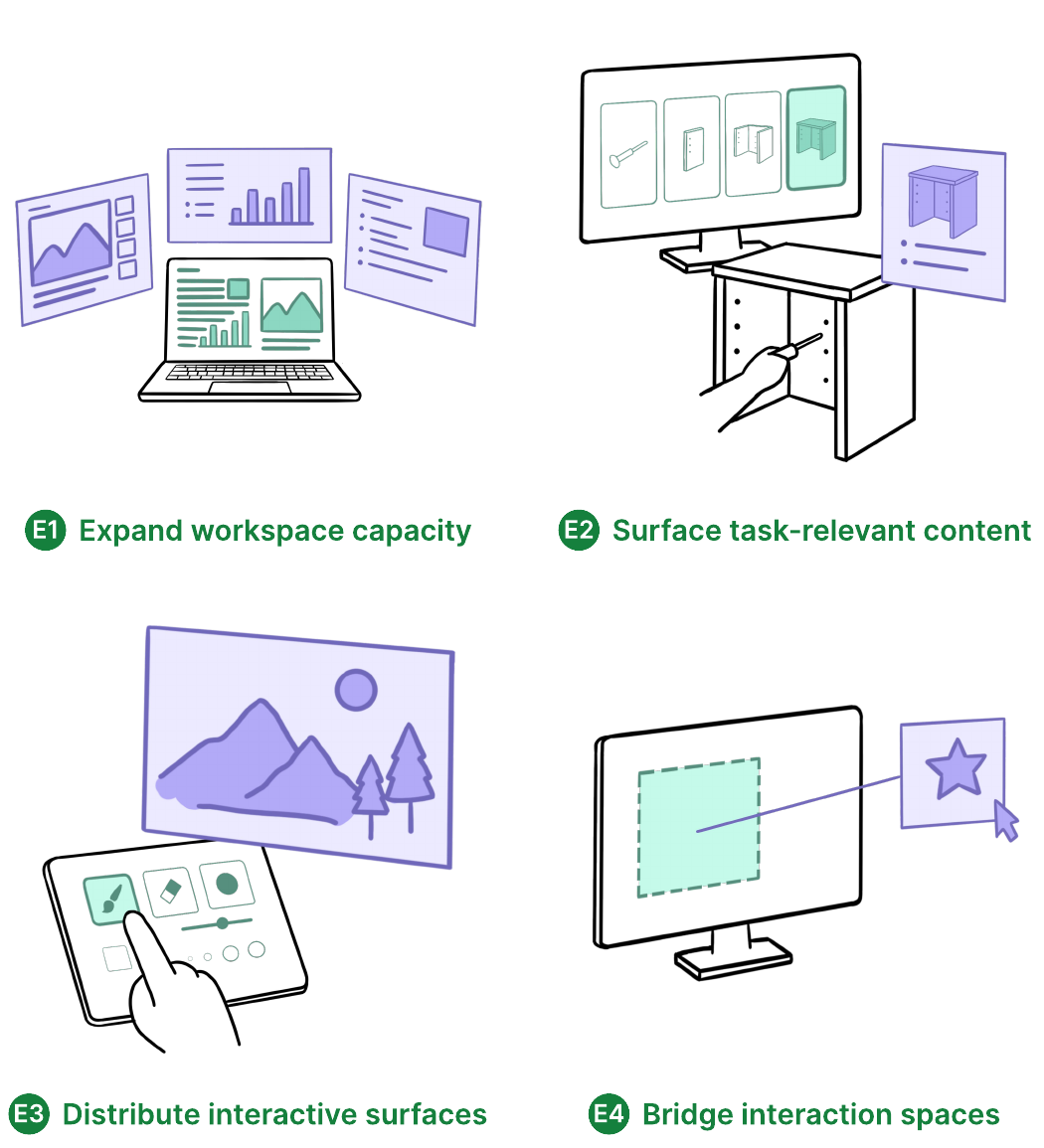}
    \caption{Illustrations of design strategies in the Extend pattern.
    (E1) Expand workspace capacity: additional virtual displays around a laptop extend the available display space beyond the physical screen.
    (E2) Surface task-relevant content: frequently consulted assembly instructions are brought closer to the user's ongoing workspace for easier access.
    (E3) Distribute interactive surfaces: drawing controls are placed on a tablet while the resulting effect is displayed on a virtual canvas in MR.
    (E4) Bridge interaction spaces: an object is dragged across the boundary between a desktop display and an adjacent MR window.}
    \Description{Four illustrations of design strategies in the Extend pattern.
    In E1, a laptop displaying charts and information is surrounded by three virtual displays containing additional visual content.
    In E2, a monitor shows a sequence of assembly steps while a user works on a physical object, with a nearby virtual display showing information about the assembled object.
    In E3, a hand interacts with drawing controls on a tablet while a landscape image is shown on a larger virtual display above it.
    In E4, a desktop monitor containing a highlighted region is connected by a drag trajectory to a nearby virtual display containing a star, with a cursor positioned beside the virtual display.}
    \label{fig:illustration-extend}
\end{figure}

\designstrategy[extendcolor]{E4}{Bridge interaction spaces} 
The system connects immersive and non-immersive spaces as adjacent or continuous interaction regions, so users can migrate content or continue an action without switching tools or interaction modes. For example, a cursor may move beyond the desktop screen into AR space~\cite{mrcd187}, an object may be dragged through a screen border into a 3D region~\cite{mrcd019,mrcd173,mrcd025}, or a brush or lasso trajectory may continue across physical and virtual surfaces~\cite{mrcd174}. The key feature is semantic continuity: although users may attend across devices, they can understand the cross-boundary action as one continuous manipulation rather than a new mapping. This continuity, however, is local to the boundary-crossing action and need not imply identical interaction semantics after the transition. Once content enters the target space, interaction may shift to its native modality or another device-specific mapping. Designers should therefore preserve the ongoing manipulation state while making such post-transition changes easy to anticipate, so users can continue the task without an abrupt interaction-mode switch~\cite{mrcd173,mrcd019,mrcd174}.

\subsubsection{\textbf{\textcolor{complementcolor}{Complement}}: Fragmented, Related, and Asymmetric}

In the Complement pattern $(n=46)$, users coordinate immersive and non-immersive devices as different but interdependent parts of the same task. The devices provide complementary representations, task stages, or interaction roles. Each device contributes task-relevant but different information or action and thus users distribute visual attention across them and need to interpret how the parts relate. Unlike Extend, Complement does not simply expand or redistribute surfaces within the same action context; it divides the task into different but connected roles. Unlike Proxy, the asymmetry here is less about action mapping and more about coordinating attention and interpretation across task views and stages.

\begin{figure}[H]
    \centering
    \includegraphics[width=0.9\linewidth]{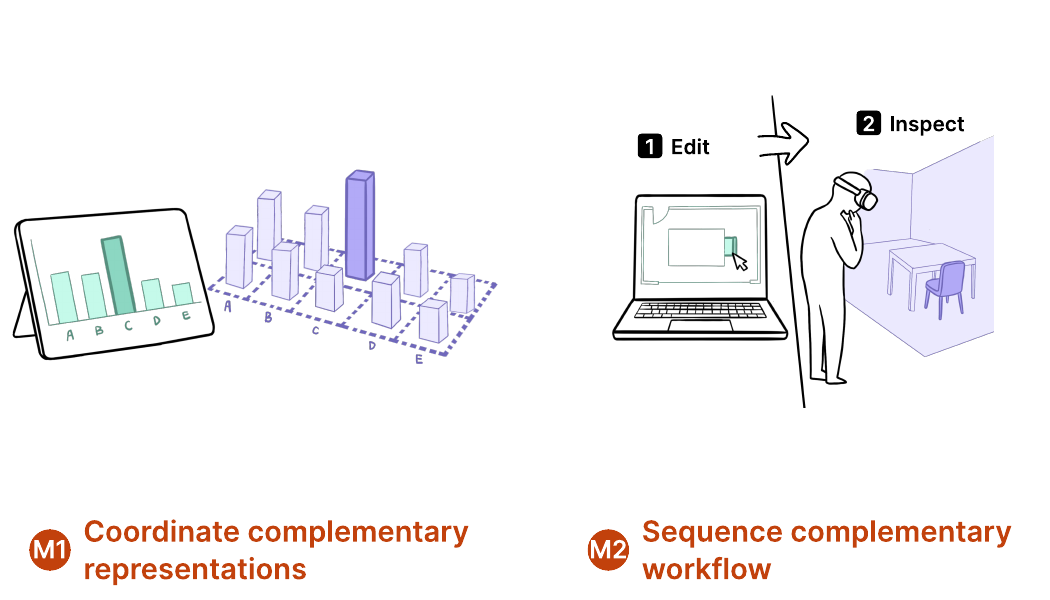}
    \caption{Illustrations of design strategies in the Complement pattern.
    (M1) Coordinate complementary representations: a tablet presents data as a 2D bar chart while MR presents the corresponding data as spatial 3D bars.
    (M2) Sequence complementary workflow: a user first edits a room layout on a laptop and then puts on the headset to inspect the resulting spatial arrangement in MR.}
    \Description{Two side-by-side line drawings. In panel (M1), a tablet displays a 2D bar chart labeled A through E alongside a corresponding set of spatial 3D bars in MR, illustrating complementary representations of the same data. In panel (M2), a user first edits a room layout on a laptop and then wears an MR headset to inspect the corresponding room arrangement in the surrounding space.}
    \label{fig:illustration-complement}
\end{figure}

\designstrategy[complementcolor]{M1}{Coordinate complementary representations} The system presents different but semantically connected representations of the same underlying object or data across devices. For example, one device may show a 2D chart, table, map, parameter panel, or overview, while the other shows a 3D spatial scene or 3D visualization, which often appears in immersive analytics, sense-making, and data visualization systems~\cite{mrcd279,mrcd028,mrcd035,mrcd169,mrcd305,mrcd071}. Some systems further couple these representations through spatial lenses, where a phone or tablet is registered as a movable window into the MR scene and displays 2D information (e.g. cross-section view) based on its position or orientation~\cite{mrcd133,mrcd255}. The design goal is to help users integrate information that cannot be fully understood from one representation alone. However, the differences between representation can create cognitive gaps. Systems bridge this through spatial continuity~\cite{mrcd028,mrcd077}, bidirectional highlighting~\cite{mrcd305,mrcd247}, synchronized feedback~\cite{mrcd133,mrcd255}, or contextual cues~\cite{mrcd246}. 

\designstrategy[complementcolor]{M2}{Sequence complementary workflow chains} The system allocates distinct but interdependent task stages across devices, forming a sequential workflow. The output of one stage becomes the input, reference, or object of manipulation for another stage. A common chain is 2D editing followed by 3D validation or editing, where the non-immersive device supports precise editing and the MR device supports in-situ inspection~\cite{mrcd003,mrcd107,mrcd124,mrcd266} or spatial adjustment~\cite{mrcd004,mrcd181,mrcd305,mrcd197,mrcd174}. Other chains include moving from 2D drawing to 3D object creation~\cite{mrcd037,mrcd207}, or from coarse spatial selection to fine-grained planar refinement~\cite{mrcd241}. Rather than transferring an entire task phase, some systems coordinate tightly coupled steps within a single interaction~\cite{mrcd241,mrcd174}. These workflows are complementary because each device supports a different form of reasoning or action within the same task. This strategy differs from the E3 since the latter usually remains part of the same immediate action context whereas this strategy supports more distinct stages of work, so users must connect representations and carry task meaning from one stage to the next.

\subsubsection{\textbf{\textcolor{standbycolor}{Standby}}: Focused and Unrelated}

In the Standby pattern $(n=6)$, users maintain focus on one device while another device remains available for a separate purpose, such as notification, status display, or awareness support. The tasks are unrelated because the standby device does not directly contribute to the user's current primary goal. One strategy is to \inlinestrategy[standbycolor]{S1}{push standby signals} into the active device, where notifications or lightweight status cues are presented within the user's current display space~\cite{mrcd011,mrcd128}. In multi-user settings, another strategy is to \inlinestrategy[standbycolor]{S2}{externalize users' status} through the standby device. Rather than serving the MR user directly, the standby device allows nearby people or bystanders to interpret the MR user's otherwise hidden state, such as facial expressions or activity-related warnings~\cite{mrcd095,mrcd288}. This supports social awareness around the MR activity while allowing the MR user to maintain focus on the primary task.

\begin{figure}
    \centering
    \includegraphics[width=0.85\linewidth]{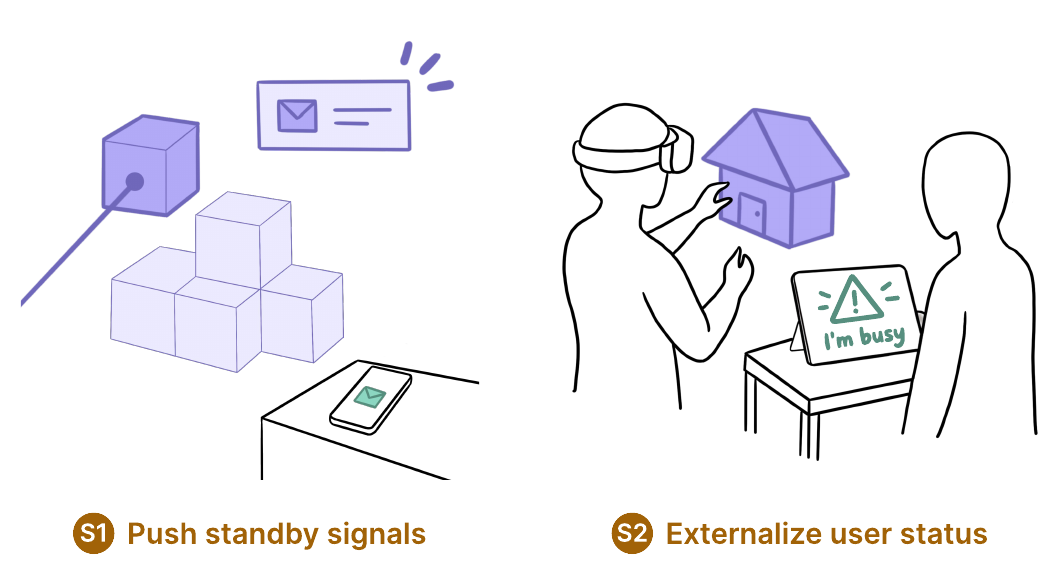}
    \caption{Illustrations of design strategies in the Standby pattern.
    (S1) Push standby signals: a notification from a smartphone is surfaced in MR while the user is engaged in an unrelated MR task.
    (S2) Externalize user status: an MR user's busy status is displayed on a nearby tablet that is not involved in the current MR task, signaling their unavailability to people nearby.}
    \Description{Two side-by-side line drawings. In panel (S1), a user works with virtual blocks in MR while a smartphone rests on a nearby table, and a message notification is surfaced in the MR view. In panel (S2), an MR user interacts with a purple virtual house while another person sees an ``I'm busy'' status on a tablet, indicating that the MR user is currently occupied.}
    \label{fig:illustration-standby}
\end{figure}

\subsubsection{\textbf{\textcolor{partitioncolor}{Partition}}: Fragmented and Unrelated}

In the Partition pattern $(n=3)$, users divide attention across devices that support separate activities, making this the primary pattern for MR cross-device multitasking. Systems can facilitate this pattern by \inlinestrategy[partitioncolor]{R1}{supporting parallel activities}. However, few of the reviewed systems were explicitly designed around this pattern. Instead, they typically enabled this through mechanisms also seen in other patterns. Some systems clone a phone screen into MR in the same way as in the Clone, but to enable potentially unrelated tasks, such as texting and photo-taking in the MR environment~\cite{mrcd154,mrcd044}. In addition, a phone may serve as a spatial lens for inspecting the surrounding physical environment~\cite{mrcd321}. This limited design focus suggests a possible gap in current MR cross-device research, which we return to in Section~\ref{subsec:research-oppor}.

\begin{figure}[h]
    \centering
    \includegraphics[width=0.4\linewidth]{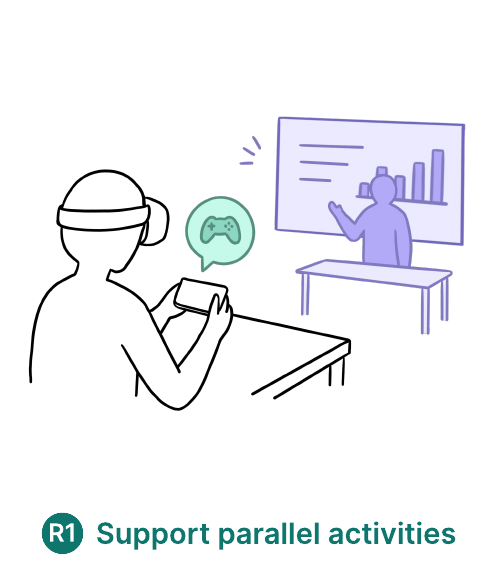}
    \caption{Illustration of the design strategy in the Partition pattern.
    (R1) Support parallel activities: a VR user follows a virtual presentation while simultaneously playing an unrelated game on a smartphone.}
    \Description{A line drawing of a seated user wearing a VR headset and holding a smartphone. A game-controller symbol beside the smartphone indicates an unrelated gaming activity, while a purple virtual presenter and presentation screen appear in VR, illustrating two parallel activities carried out across the two devices.}
    \label{fig:illustration-partition}
\end{figure}

\subsection{Combinations Across Usage Patterns} \label{subsec:combination}

\begin{figure}
    \centering
    \includegraphics[width=0.9\linewidth]{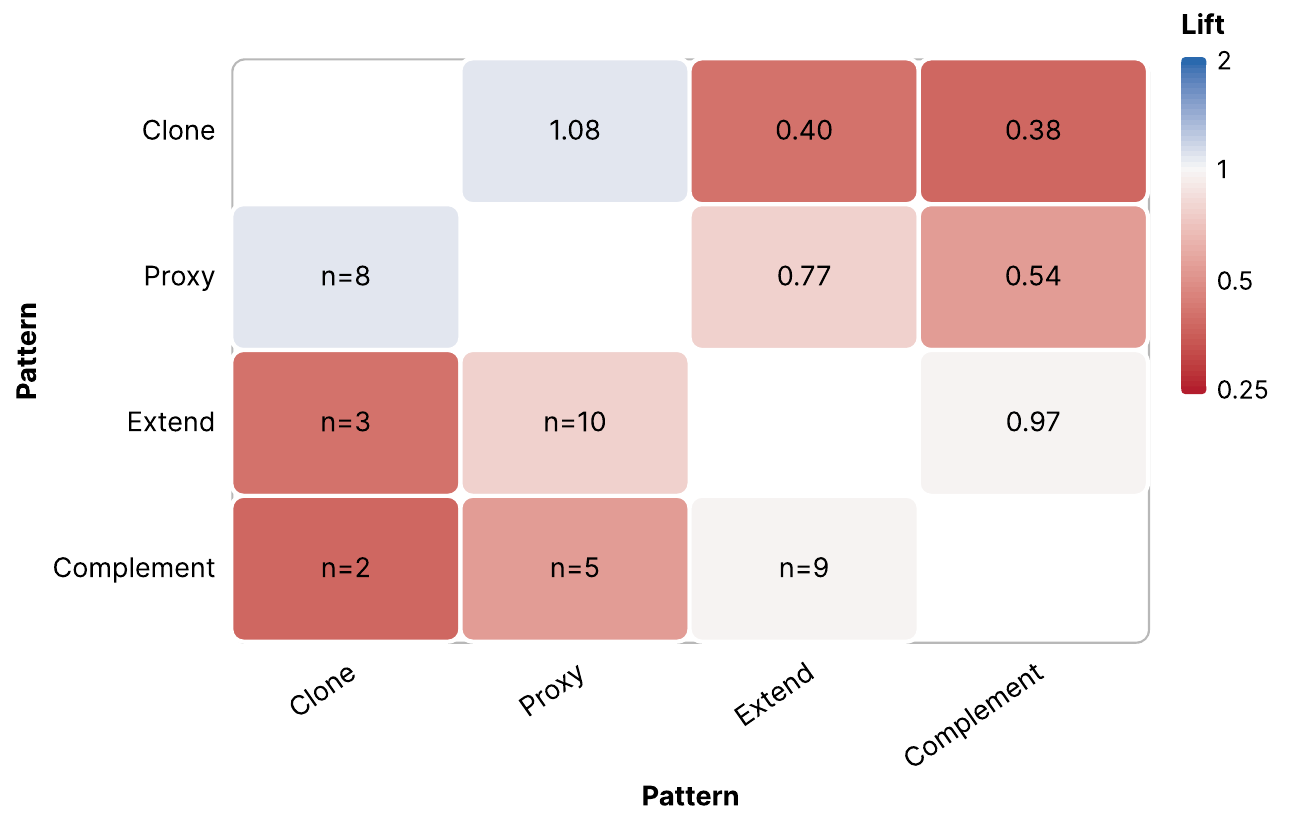}
    \caption{Pairwise combinations among the four Related usage patterns. The lower triangle shows the number of papers containing each pattern pair ($n$), and the upper triangle shows lift, representing observed relative to expected co-occurrence, with values above (below) 1 indicating more (less) co-occurrence than expected. Because a paper may contain more than two patterns, it can contribute to multiple pairwise counts; these counts are therefore not mutually exclusive.}
    \Description{A four-by-four triangular matrix showing pairwise combinations of Clone, Proxy, Extend, and Complement. Clone--Proxy occurs in 8 papers with a lift of 1.08; Clone--Extend in 3 papers with a lift of 0.40; Clone--Complement in 2 papers with a lift of 0.38; Proxy--Extend in 10 papers with a lift of 0.77; Proxy--Complement in 5 papers with a lift of 0.54; and Extend--Complement in 9 papers with a lift of 0.97. Only Clone--Proxy has a lift above 1, while Extend--Complement is close to the expected co-occurrence rate.}
    \label{fig:pattern-combination}
\end{figure}

We identified several systems that integrate multiple usage patterns. We further examined these cases to characterize how patterns were combined. 34 papers (14.3\% of the corpus) introduced systems that meaningfully combined multiple usage patterns, as shown in Fig.~\ref{fig:pattern-combination}. No particular pair showed strong systematic affinity, with lift values generally around or below 1. The three most frequent combinations were Clone--Proxy, Extend--Complement, and Proxy--Extend. Combinations involving Unrelated patterns were rare and did not form recurring pairwise structures; we therefore discuss them only in~\ref{subsubsec:less-common-combination}.

A recurring approach was to leverage Proxy as a flexible input layer within mixed-pattern designs. When used in this role, we identified two recurring design motivations: to supplement the input already supported by another pattern, or to provide an input method for the output introduced by another pattern.
To supplement existing input, Proxy is often combined with Extend, where portable non-immersive devices such as phones and tablets can simultaneously host redistributed interface components (E3) and serve as proxy controllers (X1). For example, a phone can retain familiar touch input for interfaces redistributed into XR, while dynamically transitioning into a spatial controller as application contexts change~\cite{mrcd194}. This combination retains the intuitiveness of symmetric interaction with familiar 2D interfaces while introducing specialized asymmetric controls better suited to manipulating objects in 3D MR space. Beyond altering input behaviors at the task level~\cite{mrcd194}, several systems exploit physical orientation and spatial proximity to dynamically coordinate the patterns. For example, a tablet may display redistributed UI elements when positioned vertically or flipped, while also being used as a tangible prop for selecting nearby 3D objects~\cite{mrcd052,mrcd107}. Together, these adaptive mechanism illustrate a broader coordination strategy in which task and spatial context can mediate transitions between usage patterns.

For the second motivation, Proxy is frequently paired with output-oriented strategies from Extend, Complement, and Clone. In these combinations, the other pattern determines how information is presented, such as through complementary spatial presentation (M1) or workspace expansion (E1), while Proxy provides controllers or interaction shortcuts for frequent inspection and adjustment (X1). This arrangement appears particularly useful in applications involving dense information or repeated manipulation, such as immersive sense-making, extended workspaces, and data visualization~\cite{mrcd228,mrcd026,mrcd107,mrcd234}, where relying exclusively on menus or spatial input can impose substantial interaction overhead. 
For example, pen-based interaction such as rotation and movement along the z-axis can be remapped to rotate and scale virtual slides in MR, while gaze is used for selection, reducing the need to repeatedly access explicit menus~\cite{mrcd239}. A similar division occurs when Proxy is combined with Clone: the cloned view (C1) decouples visual output from the physical non-immersive device, whereas proxy interactions preserve convenient selection and predefined shortcuts, such as taking screenshots~\cite{mrcd023}.

In Extend--Complement combinations, we identified two recurring design motivations. One is to support hybrid workflows across devices. These systems often combine a PC for dense or precise editing (M2) with a portable device that brings lightweight controls into the in-situ MR workflow (E3)~\cite{mrcd191,mrcd207}. Thus, this combination enhances users' fine-tuning ability within the complementary workflow. Extend can also bridge the asymmetric stages of the complementary workflow, supporting a smoother transition from desktop to in-situ MR interaction by preserving the active task context~\cite{mrcd191}. For example, users can construct the main program structure on a PC, then use a phone for mobile parameter adjustment and convenient text or number entry while working in MR~\cite{mrcd191}. 
The second motivation is to combine an extended 2D workspace with complementary 3D representations, a configuration commonly found in 3D design and sensemaking systems. The non-immersive surface provides familiar and precise planar interaction, while MR expands the available display space for arranging data and interfaces (E1). Complementary representations (M1) then allow the same underlying data to be examined through both symmetric and asymmetric views~\cite{mrcd028,mrcd181,mrcd037}. Their spatial continuity helps users relate the two representations and reduces the cognitive gap between them.

A similar motivation appears in Clone–Proxy combination. In multi-user settings, many systems combine a shared first-person perspective (C1) with role-specific views available only to an external collaborator (X2) to enhance common ground~\cite{mrcd111,mrcd136,mrcd088,mrcd170,mrcd177}. For example, desktop or VR analysts can access a synchronized window into one another's current view while retaining platform-specific 2D or 3D representations~\cite{mrcd111}, while tablet presenters can monitor a mirrored audience view alongside a top-down room view showing audience position and orientation~\cite{mrcd136}. Here, the same symmetric--asymmetric logic shifts from coordinating representations across devices to coordinating information across user roles, which Clone establishes a shared reference while Proxy preserves role-specific information or controls needed by individual collaborators. %

When viewed through the lens of usage dimensions, these recurring combinations reveal a more systematic structure than the pattern pairs alone suggest. Clone--Proxy and Extend--Complement both combine symmetric and asymmetric patterns while preserving the same attention regime, allowing systems to retain a stable shared representation or familiar interaction while selectively introducing device- or role-specific capabilities where they are useful for current task. Proxy--Extend additionally crosses the attention dimension. In these cases, systems often coordinate the transition between focused and distributed device use through task state or interaction need, suggesting that combinations across attention regimes require an explicit mechanism for determining when each pattern becomes appropriate. Overall, the observed combinations therefore reflect different ways of coordinating stability and specialization, as well as focused and fragmented interaction.

\subsubsection{Less Common and Incidental Combinations}
\label{subsubsec:less-common-combination}

Beyond the recurring combinations discussed above, we observed several less common or less tightly integrated forms of pattern combinations. First, several combinations appeared in only one or two systems but nevertheless exhibited a clear functional rationale. A few systems are designed for both single-user and multi-user settings, which leads to a part of Complement--Proxy~\cite{mrcd004,mrcd160}, Complement--Clone~\cite{mrcd197}, and Extend--Clone~\cite{mrcd012} combination. In single-user workflow (M2,E1), the system allows an individual user to switch between devices. By further enabling the real-time data synchronize, they can also support two users to control the different devices to collaborate around a role-specific view (X2) or shared scene (C1). These systems provide a potential direction to expand the future MR system from single usage to multi-users. Second, some systems combined multiple patterns toward the same overall task goal, but with relatively loose coupling between them. In these cases, the patterns addressed separate local needs within the task, without one pattern directly enabling or coordinating the other. Such combinations often took the form of pairing multiple input-related functions (e.g., E1+M2~\cite{mrcd174}, C2+E3~\cite{mrcd120}), combining independently useful input and output functions (e.g., C2+M1~\cite{mrcd233}), or mixing on-demand unrelated usage (R1~\cite{mrcd321}, S1~\cite{mrcd011}). Finally, additional combinations could arise if patterns outside a paper's primary design motivation were also counted. As mentioned in~\ref{subsubsec:pattern-coding}, we excluded incidental uses during pattern coding to avoid inflation. A common pattern for these cases to combine is Clone as C2 retains the basic interaction mechanism of input devices. Examples includes cloning the laptop input at extended workspace (E1)~\cite{mrcd285}, supporting basic keyboard typing or pen drawing as alternative when proposing X1~\cite{mrcd120,mrcd023} and E3~\cite{mrcd206,mrcd096}. These cases represent forms of pattern coexistence that are present in the systems but intentionally omitted from the quantitative combination analysis.

\subsection{Boundary Cases}

We encountered several boundary cases when applying either our coding framework or eligibility criteria, concerning how included systems should be classified or whether closely related systems should be included.

Some included systems used additional devices mainly as contextual sensors, such as camera-based recognition~\cite{mrcd089,mrcd116,mrcd309}, microphone or phone-call input~\cite{mrcd167}, and smartwatch heart-rate sensing~\cite{mrcd304,mrcd223,mrcd139}. Although users may not explicitly operate or attend to these devices, we retained them because their outputs directly participated in the cross-device usage episode and affected the interaction on the other device. For cognitive symmetry, we coded whether the sensed information preserved a direct semantic correspondence with its resulting effect or was transformed into a different interpretation. These cases therefore represent a more passive form of cross-device coordination rather than a separate class outside our framework.

We also encountered devices that closely resembled common consumer devices but incorporated specialized affordances, such as a tablet with a transparent display~\cite{krug2022clear}, as well as simulated devices such as virtual tablets~\cite{mrcd010}. Although their interaction concepts may transfer to more common cross-device settings, we excluded them under our existing criteria to avoid case-by-case judgments about how much specialization remained sufficiently transferable. Such systems may nevertheless help extend the framework as emerging device capabilities become more commonplace.

\section{Usage of MRxD Framework}

This section shows how MRxD can be used in practice through example system analysis, concept generation, and a formative workshop on early-stage MR cross-device design.

\subsection{Example Applications}

\paragraph{Analyzing existing systems.}
MaRginalia supports classroom note-taking by projecting a tablet interface into MR to reduce visual attention shifts between the tablet and the lecture~\cite{mrcd023}. Its cloned tablet screen and familiar pen-based note-taking fit the \textbf{Clone} pattern, while pen interactions that map the pen to an MR cursor can be analyzed as \textbf{Proxy}. This pattern analysis clarifies the system's rationale: its interaction techniques consistently support focused attention. 
By mapping Proxy designs into MRxD, we further identify a latent trade-off, where some Proxy mappings remain inconsistent, such as using pen-down for button activation but pen squeezing for manual snapshot capture. Although the paper does not frame this as a mapping issue, a participant's workaround of capturing transcripts by tapping a dot suggests that inconsistent asymmetric mappings may introduce local friction.

\paragraph{Generating new design concepts.}
Consider a desktop-to-MR generative design workflow where a designer edits prompts and parameters on a desktop and inspects generated 3D objects in MR. MRxD characterizes this workflow as \textbf{Complement}: both devices support the same authoring task, but attention is fragmented between desktop editing and MR inspection, and the relationship is cognitively asymmetric because text prompts and parameters differ from 3D outputs. This diagnosis suggests bridging strategies that make the asymmetry easier to interpret. For example, hovering over a prompt phrase or parameter on the desktop could highlight the corresponding object component or material in MR. The workflow could also introduce \textbf{Proxy} for quick inspection, such as using a phone's IMU to manipulate the MR object through lightweight motion input.

\subsection{Formative Design Workshop} \label{subsec:workshop}

To understand how designers apply MRxD in practice and collect feedback, we conducted a formative one-on-one design workshop with six HCI designers (2 female, 4 male; ages 23--29), all with AR/VR-related design or research experience, including four academic researchers and two industry engineers. The workshop adapted three stages from the Google design sprint~\cite{knapp2016sprint}---map, diverge, and decide---to support problem framing, design exploration, and reflective selection. In the \textbf{map} stage, participants chose whichever of the two design cases they felt would better support their ideation and spent about five minutes identifying user needs and cross-device design challenges: VR classroom learning with a headset, tablet, pen, and laptop ($n=2$), or AR home decoration with an AR headset, smartwatch, smartphone, and desktop ($n=4$). The cases provided concrete design contexts and were not intended for comparison. In the \textbf{diverge} stage, participants first spent ten minutes designing from their own knowledge while thinking aloud and sketching. We then introduced MRxD, including its three dimensions, six usage patterns, and related design strategies, after which participants spent another ten minutes revisiting and extending their ideas with the framework when appropriate. In the \textbf{decide} stage, participants spent about ten minutes refining their ideas and reflecting on coordination trade-offs using MRxD. The workshop ended with a 15-minute semi-structured interview on MRxD's usefulness and other feedback.
We collected participants' initial and revised design artifacts, session notes, and interview transcripts. We analyzed these materials using reflexive thematic analysis~\cite{Braun01012006}. The study was approved by the local IRB. %

\subsubsection{Results}

Our results show that participants used MRxD in three main ways.
\textbf{Revealing overlooked needs and design blind spots.} After MRxD was introduced, participants added elements that were absent from their initial designs, including multi-user support (P1, P2, P5, P6) and missing input or output components (P1, P3, P4, P6). Participants described the design strategies as directly useful (P2--P4), and P1 described usage patterns as a ``connector between usage behavior and design strategies'' that ``helped avoid redundant elements and find supplemental designs.'' 

\textbf{Encouraging capability-level decomposition of devices.} Participants initially tended to assign whole devices to broad roles, such as computer for editing, AR for display, or phone for note-taking. After using MRxD, several decomposed devices into capabilities such as input, output, sensing, and transfer. For example, P1 and P3 moved toward reusing keyboard input inside VR, while P6 shifted from treating the phone as a note-taking tool to using it as a proxy controller for placing furniture in MR. 

\textbf{Prompting dimension-guided design and reasoning.} The three dimensions helped participants reason about coordination choices. User Attention led participants to consider whether users should remain focused in MR or move across devices, as when P6 proposed cloning phone notes onto an AR virtual screen to reduce repeated glances at the phone. Cognitive Symmetry prompted participants to move beyond direct or mirrored device relationships and explore asymmetric mappings, such as using a pen (P1) or smartphone (P4) as a laser pointer for selection. P4 and P5 further suggested that this dimension-guided reasoning could be better supported through a more interactive format, such as a website that allows designers to browse usage patterns, compare dimensions, and access relevant examples during ideation.

\section{Discussion} \label{sec:discussion}

\subsection{MRxD as a Design Scaffold}

MRxD supports design reasoning by connecting intended user activities with concrete choices about cross-device coordination. The formative workshop illustrates how designers used this connection in early-stage design. Participants used the framework to identify support that was absent from their initial concepts, reconsider broad device-level roles in terms of specific capabilities such as input and output, and reflect on coordination choices through the usage dimensions.
These observations suggest two design roles of MRxD. First, the framework provides an analytic structure for examining whether a proposed design aligns with its intended design goal. For example, a design intended to maintain an immersive and focused experience may still introduce interactions that require users to visually attend to a secondary device. Mapping the design back onto the usage dimensions makes such tensions explicit and can prompt designers to reconsider the design. This highlights the value of the framework in making design reasoning explicit, allowing underlying assumptions, trade-offs, and inconsistencies to be more readily examined and revised~\cite{maclean1991questions}. In addition, patterns and strategies not represented in the current design can similarly suggest alternative design possibilities. When multiple patterns appear within a workflow, MRxD can further help designers reason about how they interact and what coordination trade-offs their combination may introduce (Section~\ref{subsec:combination}).
Second, MRxD can serve as a generative scaffold that connects intended usage behavior to concrete design strategies. Designers can begin with a usage behavior, identify the corresponding pattern, and use its associated strategies as starting points for design exploration. Such structured prompts may be particularly useful for less-experienced designers by helping them consider alternatives beyond their initial concepts~\cite{yilmaz2010cognitive,bornoe2016facilitating}.
In the future, because these patterns are defined by usage rather than specific device configurations, MRxD may also accommodate emerging devices and interaction techniques according to the cross-device relationships they create, with recurring new realizations further extending the strategy set. This allows MRxD to remain useful as a design scaffold even as MR devices and interaction techniques evolve.

\subsection{Designing Along MRxD Dimensions} \label{subsec:design-along-dimension}

For systematic coding across a large corpus, we operationalized the three MRxD dimensions as two-sided distinctions. In design use, however, these dimensions can also be treated as spectra, allowing designers to steer a system toward more focused or fragmented attention, or toward more symmetric or asymmetric cognition, by implementing specific designs. This also allows the analyze of user behavior along the spectra, leading to more detailed design for the middle state. 

\subsubsection{Shifting User Attention}

As discussed in Section~\ref{subsubsec:attention}, fragmented attention often resulted from distributed resources or controls that required visual access across devices. To shift user attention toward the focused end, strategies in Clone and Proxy suggest a common principle: \emph{reduce the need to visually confirm the secondary device}. Continuous input and stable, learnable, or tactile controls can support this by allowing users to act while monitoring feedback in MR, whereas visually locating controls or confirming actions (e.g. E3, M2) draws attention back to the device. This is consistent with prior work showing lower visual demands for gestural than soft-button interaction~\cite{bragdon2011experimental}. 
Our multi-pattern results further reveal input--output decoupling as another way to support focused attention. Designers can separate visual output from the physical device while retaining that device for proxy input, combining Clone and Proxy strategies to keep visual attention in MR. For example, a tablet can serve as a trackpad or stable input surface while the corresponding cursor or guidance appears near the MR task~\cite{mrcd023}. When visual access across devices is unavoidable, as often occurs in Extend, Complement, and Partition, designers can instead manage fragmentation by reducing cost of attention switches. AR resumption cues can be used as contextual guidance that facilitates returning to an interrupted task~\cite{bahnsen2024augmented}. The workflow also can be optimized to shift the attention closer to the focused end by reducing the switch frequency.

Attention can also shift along the spectrum as user state or momentary task demands change, even within the same design. One source is familiarity: users may initially inspect a device to locate controls, but require less visual guidance as location memory and motor control develop with experience~\cite{jiang2022learning}. Systems could therefore provide stronger cross-device cues during early use and gradually reduce this scaffolding. Attention may also shift temporarily when a task requires greater visual precision, such as locating an infrequently used control or performing a precision-critical action. These foreseeable shifts suggest that systems can adapt their support accordingly, providing visual guidance when confirmation is needed while otherwise allowing attention to remain focused on MR.

\subsubsection{Managing Cognitive Symmetry}

At the system level, designers can build on the bridging mechanisms synthesized from our corpus to manage Cognitive Symmetry. In particular, cues used for system alignment between representations can also make asymmetric interaction mappings more interpretable. For example, when different touch regions on a Proxy device are mapped to distinct MR commands~\cite{mrcd107}, labels or icons placed at these regions can make the mapping directly observable. However, when such cues require users to inspect another device or representation, reducing the cognitive gap may introduce fragmented attention, revealing a potential trade-off between usage dimensions. 

Designers can also reduce cognitive gaps by preserving continuity across states and grounding new mappings in familiar schemas. SeamlessVR showed that continuously transforming an immersive visualization into its corresponding screen view reduced the cognitive effort required to connect the two representations~\cite{mrcd186}. A similar principle appears in our observations, where Extend bridges asymmetric stages of a Complement workflow by preserving the active task context as users transition from desktop to in-situ MR interaction~\cite{mrcd191}, echoing broader work on transitional interfaces. Familiarity provides another form of bridging: consistent with Structure-Mapping Theory~\cite{theory-smt}, designers can favor mappings that preserve familiar relational schemas, such as treating a phone as a controller~\cite{mrcd194} or a pen as a pointer~\cite{mrcd055}. More specialized mappings that depart from familiar device roles~\cite{mrcd107,mrcd052} may instead require stronger feedback or guidance while users form the corresponding mental model.

Our E4 findings further suggest that Cognitive Symmetry can be local to an interaction state rather than stable throughout a workflow. A cross-boundary action may remain directly interpretable during transfer, yet a subsequent interaction may become asymmetric once this interaction ends and adopts the target space's device-specific mapping~\cite{mrcd173,mrcd019,mrcd174}. Designers should therefore consider how easily users can carry their interpretation across such state transitions, not only the symmetry of each state in isolation. Similar shifts can arise in adaptive systems that redistribute interfaces based on contextual signals such as gaze or distance~\cite{mrcd005,mrcd149}. Such adaptation may improve contextual fit, but users must understand both the resulting state and why the correspondence changed; prior work similarly shows that unpredictable interface adaptations can cause confusion and relearning costs~\cite{todi2021adapting}. Designers should therefore consider not only the Cognitive Symmetry of each state, but also whether users can carry their interpretation across transitions between them.

\subsection{Research Opportunities} \label{subsec:research-oppor}

Unrelated patterns remain underexplored in current MR cross-device research, partly due to our search strategy and the field's focus on primary MR tasks. Nevertheless, unrelated usage remains important because users may still need to handle secondary activities while working in MR. Future work could explore bridging mechanisms that help users remain aware of activities across devices and smoothly switch between related and unrelated tasks. For example, prior systems have used visual focus to infer whether users are attending to an unrelated phone task or using the phone as an MR controller~\cite{mrcd011}. Future systems could extend this direction through activity awareness, visual cues, and adaptive reminders that support smoother suspension and resumption.

Another opportunity is to further examine interaction flows from MR to non-immersive devices. Echoing the MR-HUI review~\cite{mr-hui}, most MR cross-device systems focus on integrating non-immersive devices into MR, while fewer systems explore the reverse direction. Existing examples include gesture or controller-as-mouse interactions for non-immersive devices~\cite{mrcd031,mrcd025,mrcd013}, data migration across device boundaries~\cite{mrcd019}, and tablet-based fish-tank views for inspecting 3D content~\cite{mrcd174}. Future work could more systematically explore this direction, for example, to consider how MR input modalities can serve as proxy inputs for non-immersive devices. Existing MR interaction vocabularies, such as gaze/gesture targeting~\cite{gaze-input}, palm-based interfaces~\cite{palm-based}, and mid-air gesture sets~\cite{mid-air-gesture}, may provide a basis for such mappings, but they have rarely been examined as MR-to-non-immersive Proxy strategies.

MRxD also suggests the need for configurable design toolkits. Many strategies contain recombinable variants, such as input sources, output targets, mapping rules, and bridging cues, yet current systems often implement them as one-off designs. Inspired by prior cross-reality toolkits~\cite{vr-reception}, an MRxD-based toolkit could help designers configure these components and examine how a design behaves across the framework dimensions.

\subsection{Limitations}

We identify several limitations of this study. First, our search strategy and review scope may not capture the full landscape of MR cross-device research. Despite accounting for fragmented terminology and including device-specific keywords, relevant work may still be missed when its cross-device characteristics are not reflected in searchable metadata (e.g., Friedl-Knirsch et al.~\cite{friedl-knirsch2024collaborative}). In addition, our focus on HMD-based MR with common non-immersive devices excludes adjacent settings such as handheld MR, spatial AR, and systems using specialized devices. Our exclusion of papers of four pages or fewer may omit relevant short-form contributions. MRxD should therefore not be interpreted as an account of all MR cross-device systems or design strategies, and future work could examine how the framework extends to these settings.

Second, developing the MRxD dimensions involved interpretive analytical decisions and is therefore not fully reproducible in the sense that another research team, or a different initial development sample, might derive a different but plausible set of dimensions from the same corpus. The initial 20\% development corpus may also have made some dimensions more salient and thereby shaped the subsequent analytical trajectory. We sought to mitigate this dependence by treating the initial framework as provisional rather than fixed and repeatedly testing and refining the dimensions against the remaining corpus. MRxD therefore should not be interpreted as the uniquely optimal decomposition of the design space, but as an empirically grounded and parsimonious abstraction developed to support the analytical goals of this study.

Third, MRxD operationalizes each usage dimension as a binary distinction, although the underlying phenomena may lie on a spectrum or change over time. This abstraction supports parsimonious comparison across the corpus but may compress intermediate cases and corresponding design strategies. %

Fourth, our analysis retained usage episodes central to each paper's design motivation or contribution rather than exhaustively coding all supported interactions. This focuses the framework on patterns actively explored in prior work, but makes episode selection partly dependent on authors' framing and our interpretation. Accordingly, pattern distributions reflect research emphasis rather than behavioral prevalence, and secondary patterns or combinations may be underrepresented. Future work could systematically examine all instantiated episodes within systems or conduct in-the-wild data collection to provide a more comprehensive account of pattern co-occurrence and prevalence.

\section{Conclusion}
This paper contributes MRxD, a usage-pattern framework for MR cross-device interaction. Based on a systematic review of 238 papers, MRxD characterizes usage episodes through User Attention, Task Relatedness, and Cognitive Symmetry, and identifies six recurring patterns with associated design strategies. A formative design workshop further suggests how the framework can support design reflection and concept generation. This paper offers MR community valuable insights for analyzing and designing cross-device systems that coordinate HMDs with non-immersive devices.

\bibliographystyle{ACM-Reference-Format}
\bibliography{references}

\appendix

\section{Search Query}
\label{app:search-query}

The search combined two sets of terms capturing cross-device systems with a set of terms describing immersive reality technologies. A small number of terms that inherently refer to hybrid or cross-reality interfaces were searched independently and therefore were not combined with the Reality Set. In abstract form, the query was:

$$
\texttt{((Main Set OR Supplementary Set) AND Reality Set) OR Standalone Set}.
$$

\subsection{Main Set.}
The primary cross-device search terms were:

\texttt{"Hybrid Virtual Environment" OR
"Hybrid Interaction" OR
"Cross-Device" OR
"Cross Device" OR
"Cross-Surface" OR
"Cross Surface" OR
"Multi-Device" OR
"Multi Device" OR
"Multi-Display" OR
"Multi Display" OR
"Distributed User Interface" OR
"Distributed User Interfaces" OR
"Transitional Interface" OR
"Transitional Interfaces"}

\paragraph{Standalone Terms.}
The following terms inherently describe hybrid or cross-reality interfaces and were therefore searched independently, without being combined with the Reality Set:

\texttt{"Hybrid User Interface" OR
"Hybrid User Interfaces" OR
"Complementary Interface" OR
"Complementary Interfaces" OR
"Augmented Display" OR
"Augmented Displays" OR
"Cross-Reality" OR
"Cross Reality"}

\subsection{Supplementary Set.}
To capture studies that described cross-device configurations primarily through the devices or interaction arrangements involved, we additionally searched:

\texttt{"phone" OR "phones" OR
"smartphone" OR "smartphones" OR
"tablet" OR "tablets" OR
"tabletop" OR "tabletops" OR
"laptop" OR "laptops" OR
"desktop" OR "desktops" OR
"smartwatch" OR "smartwatches" OR
"smart watch" OR "smart watches" OR
"keyboard" OR "keyboards" OR
"mouse" OR "mice" OR
"pen" OR "pens" OR
"pencil" OR "pencils" OR
"stylus" OR "styluses" OR
"asymmetric interaction" OR
"asymmetric interactions" OR
"asymmetrical interaction" OR
"asymmetrical interactions" OR
"companion device" OR
"companion devices"}

\subsection{Reality Set.}
Both the Main Set and Supplementary Set were combined with the following reality-related terms:

\texttt{"Augmented Reality" OR "AR" OR
"Extended Reality" OR "XR" OR
"Mixed Reality" OR "MR" OR
"Virtual Reality" OR "VR"}

\end{document}